\documentclass[manuscript, 10pt, screen, %nonacm %authorversion
]{acmart}

\usepackage[utf8]{inputenc}

\usepackage{comment} % \begin{comment} \end{comment}
\usepackage{graphicx}
\usepackage{float}
\usepackage{sidecap}
\usepackage{amsmath}
\usepackage{multirow} % Tables merged cells
\usepackage{tabularx} % \begin{tabularx} \end{tabularx}
\usepackage{booktabs} % Tables \toprule, \midrule, \bottomrule
\usepackage{xspace} % \xspace

\usepackage{hyperref} % Supports links in \url{} and internal pdf links
\hypersetup{
	breaklinks=true,
	colorlinks=true,
}
\graphicspath{{./}{./figures/}}

\usepackage[shortcuts]{glossaries-extra}
\glsdisablehyper

\newabbr{defi}{DeFi}{decentralized finance}
\newabbr{adefi}{aDeFi}{anonymous decentralized finance}
\newabbr{dapp}{dApp}{decentralized application}
\newabbr{sc}{SC}{smart contract}
\newabbr{dao}{DAO}{decentralized autonomous organization}

\newabbr{mev}{MEV}{maximal extractable value}
\newabbr{dlt}{DLT}{distributed ledger technology}
\newabbr{tvl}{TVL}{total value locked}
\newabbr{dact}{DACT}{decentralized anonymous agnostic cross-chain transfer}
\newabbr{poc}{PoC}{proof of concept}
\newabbr{dap}{DAP}{decentralized anonymous payment scheme}
\newabbr{evm}{EVM}{Ethereum virtual machine}
\newabbr{eth}{ETH}{Ethereum native cryptocurrency} 
\newabbr{matic}{MATIC}{Polygon native cryptocurrency} 
\newabbr{erc20}{ERC20}{Cryptocurrency commonly known as Token that follows specific standards. It is created at an application level on the Ethereum blockchain through SCs} 
\newabbr{js}{JS}{JavaScript} 
\newabbr{api}{API}{application protocol interface} 
\newabbr{id}{ID}{Identifier} 
\newabbr{mimc}{MiMC}{minimal multiplicative complexity}
\newabbr{ide}{IDE}{integrated development environment} 

\newabbr{ecdsa}{ECDSA}{elliptic curve digital signature algorithm} 
\newabbr{mpc}{MPC}{multi party computation} 
\newabbr{snark}{SNARK}{succinct non-interactive argument of knowledge}
\newabbr{zkp}{ZKP}{zero knowledge proof}
\newabbr{zksnark}{zk-SNARK}{zero knowledge SNARK} 

\newcommand{\redefineacronyms}{%
    \glsresetall
}

\newcommand{\proto}{DACT\xspace}
\newcommand{\infra}{SurferMonkey\xspace}
\newcommand{\tornado}{Tornado Cash\xspace}

\usepackage[%nameinlink
,capitalize]{cleveref}
\renewcommand{\Cref}[1]{\cref{#1}} % always have short forms
\renewcommand{\autoref}[1]{\cref{#1}} % avoid hyperref scheme

\setcopyright{rightsretained} % No copyright notice required for submissions
\acmConference[~]{~}{~}{~}
\acmYear{2022}
\date{}

\title{SurferMonkey: A Decentralized Anonymous Blockchain Intercommunication System via Zero Knowledge Proofs}

\author{Miguel D\'{i}az Montiel}
    \affiliation{\institution{\'{E}cole Polytechnique F\'{e}d\'{e}rale de Lausanne (EPFL)} \country{Switzerland}}
    \affiliation{\institution{Universitat Polit\`{e}cnica de Catalunya (UPC)} \country{Spain}}
    \email{miguel.diaz.montiel@estudiantat.upc.edu}
\author{Rachid Guerraoui}
    \orcid{0000-0002-4794-8902}
    \affiliation{\institution{\'{E}cole Polytechnique F\'{e}d\'{e}rale de Lausanne (EPFL)} \country{Switzerland}}
    \email{rachid.guerraoui@epfl.ch}
\author{Pierre-Louis Roman}
    \orcid{0000-0001-5741-1490}
    \affiliation{\institution{\'{E}cole Polytechnique F\'{e}d\'{e}rale de Lausanne (EPFL)} \country{Switzerland}}
    \email{pierre-louis.roman@epfl.ch}

\thanks{This report is based on the Master's thesis of Miguel D\'{i}az Montiel defended on July 13, 2022 at EPFL.\\
Authors’ addresses: Miguel D\'{i}az Montiel, miguel.diaz.montiel@estudiantat.upc.edu; Rachid Guerraoui, rachid.guerraoui@epfl.ch; Pierre-Louis Roman, pierre-louis.roman@epfl.ch}

\begin{document}

\redefineacronyms %%% Redefine acronyms before and after abstract
\begin{abstract}
    
Blockchain intercommunication systems enable the exchanges of messages between blockchains.
This interoperability promotes innovation, unlocks liquidity and access to assets.
As of March 2022, the \ac{tvl} in these systems was of \$21.8 billion~\cite{coindesk}.
However, blockchains are isolated systems that originally were not designed for interoperability.
This makes cross-chain communication, or bridges for short, insecure by nature. 
More precisely, cross-chain systems face security challenges in terms of selfish rational players such as \ac{mev} and censorship. 
As of July 2022, the top 3 bridge hacks account for more than \$1.5 billion in losses~\cite{rekt} and the aggregated value extracted from the users using \ac{mev} techniques is \$663 million~\cite{flashbots}. 

We propose to solve these challenges using \acp{zkp} for cross-chain communication.
Securing cross-chain communication is remarkably more complex than securing single-chain events as such a system must preserve user security against both on- and off-chain analysis.

To achieve this goal, we propose the following pair of contributions: the \proto protocol and the \infra infrastructure that supports the \proto protocol.
The \ac{dact} protocol is a global solution for the anonymity and security challenges of agnostic blockchain intercommunication. 
\proto breaks on- and off-chain analysis thanks to the use of \acp{zkp}.
\infra is a decentralized infrastructure that implements \proto in practice.
Since \infra works at the blockchain application layer, any decentralized application (dApp) can use \infra to send any type of message to a dApp on another blockchain.
With \infra, users can neither be censored nor be exposed to \ac{mev}. 
By applying decentralized proactive security, we obtain resilience against selfish rational players, and raise the security bar against cyberattacks. 
We have implemented a \ac{poc} of \infra by reverse engineering \tornado and by applying IDEN3 \ac{zkp} circuits. 
\infra enables new usecases, ranging from anonymous voting and gaming, to a new phase of \ac{adefi}.

\end{abstract}

\maketitle

\clearpage
%%% Remove colored links in the table of contents
{
    \hypersetup{linkcolor=black}
    \tableofcontents
}

\clearpage
\redefineacronyms %%% Redefine acronyms before and after abstract

\section{Introduction}
\label{sec:introduction}

The industry has developed multiple blockchain systems to tackle different problems. As more blockchains systems began to appear on the market, the need for a cross-blockchain communication system became evident. This led to the development of the so called blockchain bridges.%
\footnote{In the industry, a blockchain bridge is commonly reserved for applications that solely make value transfers from one chain to another and that have their own liquidity pools. However, the nomenclature of cross-chain messaging system is reserved for protocols that \acp{dapp} use to transfer any type of message. For simplicity, we are addressing ``blockchain bridges'' as equal as ``cross-chain messaging systems''.}
A blockchain bridge is just a message passing system between different blockchains. As more blockchains communicate with each other, more value can be accessed. This improves the liquidity from the assets and it brings new projects into the table, ranging from data services, gaming, to \ac{defi}.

\begin{figure}[H]
\centering
  \includegraphics[width=0.8\textwidth]{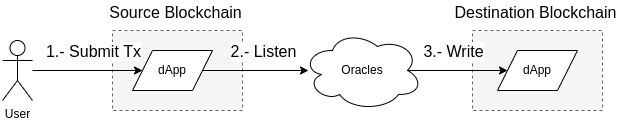}
  \caption{Traditional cross-chain communication protocol.}
  \label{fig:Bridge}
\end{figure}

A typical cross-chain communication system is depicted in \cref{fig:Bridge}. First, the user submits a transaction request into the source blockchain where it calls the dApp. Second, oracle nodes listen in to the events from the dApp and locally transform the source blockchain event into a destination blockchain transaction. Third, the oracles submit the transaction into the destination blockchain.

\subsection{Cross-Chain Ecosystem Drawbacks}
The interconnection of different blockchain systems brings new problems into the industry (game theory and censorship), inherits the drawbacks from a single blockchain ecosystem (privacy), and maximizes pitfalls from the isolated blockchain environments (\ac{mev}).

Through game theory standards, the actors of a system are seen as rational players that always look for their best interests. Censorship is the ability that an entity has to silence a user voice or transaction. MEV is a rational permissionless move where the players extract as much as possible value from their rivals without requiring their consent. Finally, in a blockchain system, users cannot have privacy nor anonymity, therefore this data can be used as an attack vector.

\subsubsection{First problem: game theory}
The new problem that arises with cross-blockchain communication systems is: How to obtain cross-chain message replication on environments that weren't designed to be interconnected? To solve this, the industry created a set of actors called oracles. The oracles hold the cryptography keys to input data on the destination blockchains. The oracles are machines run by persons or entities that, simply put, listen to the events that happened on blockchain A, and submit those events into blockchain B.
The solution brings a new variable into the equation: How can we trust the oracles?

All together, the oracles are the new entities that store the keys to secure the cross-chain messaging system. However, oracles are subject to bribery, rational thinking, and cyberattacks.

\subsubsection{Second problem: \acs{mev}}
The \acf{mev} gaming vector is performed by having nodes change the order of transactions executed. For example, an MEV player can frontrun or sandwich other transactions~\cite{sandwich}. An MEV player relies on knowing the negotiation parameters from a rival, to extract as much value from the rival as possible, before the rival transaction settles. This is possible, because the MEV players know beforehand how much more a specific rival is willing to pay for a certain asset. The MEV players can be the blockchain miners, oracles or another user.

\begin{figure}[tbh]
\centering
  \includegraphics[width=0.5\textwidth]{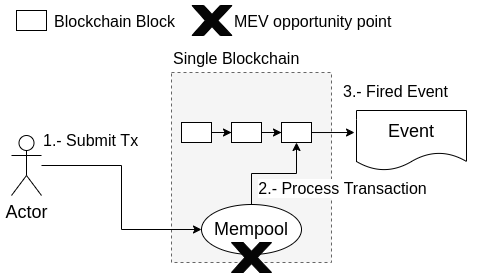}
  \caption{MEV Single-Chain opportunity points.}
  \label{fig:mevpointsSingle}
%\end{figure}

%\begin{figure}[H]
%\centering
    \vspace{2\baselineskip}
  \includegraphics[width=0.99\textwidth]{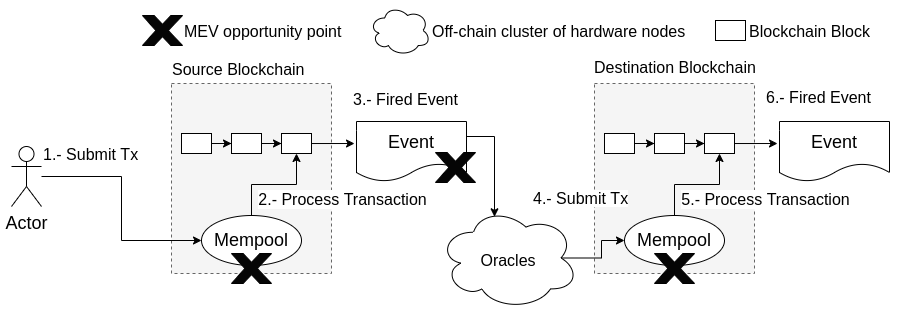}
  \caption{MEV Multi-Chain opportunity points.}
  \label{fig:mevpointsMulti}
\end{figure}

In an isolated blockchain scenario we have a single MEV point: the blockchain mempool (\cref{fig:mevpointsSingle}). While on a cross-chain communication system we have three MEV opportunities: the source chain mempool, the source fired event, and the destination chain mempool (\cref{fig:mevpointsMulti}). In a cross-chain system, the MEV players have more time to orchestrate a move prior to the rival transaction settlement. The MEV players can go even further and, if the business logic allows it, they can potentially extract value on each step of the cross-chain system.

\subsubsection{Third problem: privacy}
By default blockchains have a public register (ledger) of all the actions that happened, as well as the deeds that are about to happen (mempool). Both ledger and mempool are transparent. Therefore the whole user history can be traced. This has an impact on the privacy of the users as this data can be exploited as an attack vector and can negatively impact an individual. For example, maybe a user does not want to be publicly exposed while casting a vote within a Decentralized Autonomous Organization (DAO). 

Furthermore, in a cross-chain ecosystem, we can also trace not only the on-chain data, but also the off-chain flow. Which means that we can know that Alice transferred a message from blockchain A to blockchain B, and performed certain computational steps on blockchain B. We are able to trace the off-chain data from the cross-communication system because the user intent is transparent on the source chain. Even if that intent is obfuscated in the source chain, the oracles can still trace the off-chain footprint as they know which specific package to push from blockchain A to blockchain B which breaks the users privacy. 

All in all, transparent data on and off chain can be mined. This data can be used to train specific machine learning models in order to weaponize artificial intelligence.

\subsubsection{Fourth problem: censorship}
As we move from single blockchain environments into the multi-chain ecosystems, we find ourselves with censorship problems by the oracles against the users, dApps, and blockchains. 
The oracles can see a specific user request and destiny to perform a cross-chain transfer; the oracles have the power to silence this specific user by rejecting its cross-chain transfer requests. 
The oracles can also decide to silence the dApps. The data is transparent and the oracles can see from which dApp the cross-chain transactions originated and where is it heading to.

Furthermore, even if the cross-chain data is obfuscated, the oracles still have to push packages from the source chain into the destination chain. This off-chain footprint allows for the oracles to know each package flow. With this, the oracles can reject packages coming from blockchain A, or packages with blockchain B as a destiny. The oracles can even isolate a specific communication path, for example from blockchain B to blockchain C.

The oracles motivations to censor the network can range from maximizing profit from a DeFi project, silencing a user vote, to making a riot against public or private institutions.

\subsection{Solution: \proto and \infra}

To solve these four problems, we propose two contributions. First, the \proto protocol which is based on Proofs of Membership via ZKPs. Second, the \infra decentralized infrastructure that supports the \proto protocol. The \infra system has been built up to provide anonymity to the users while raising up the cyberattacks security threshold. We have designed \infra with four goals in mind: rational players resistance, MEV resistance, privacy and censorship resistance.

\begin{figure}[tbh]
\centering
  \includegraphics[width=0.99\textwidth]{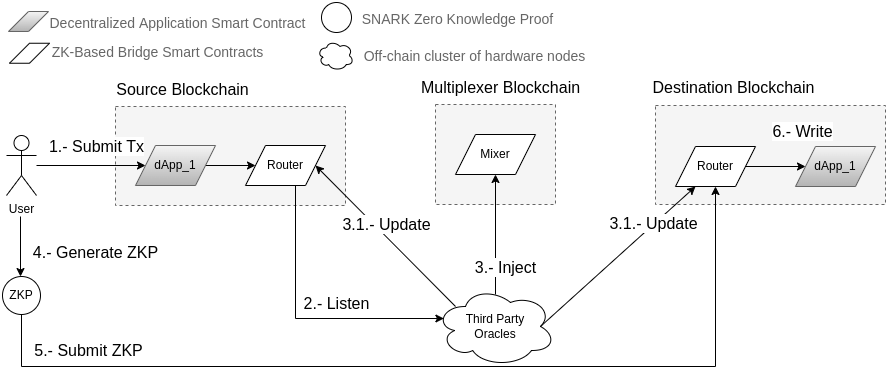}
  \caption{\infra vision of a zero knowledge blockchain intercommunication protocol.}
  \label{fig:zkBridgeVision}
\end{figure}

\infra opens new secure ways for innovation. Developers can introduce new cross-chain dApps where the payload data must be obfuscated during transit. For example, \infra allows private cross-chain voting for Decentralized Autonomous Organizations (DAOs). Furthermore, we introduce the concept of applied anonymous Decentralized Finance (aDeFi) where users can remain private while performing swaps, loans, and trades. A simplified \infra overview vision, based on Zero Knowledge for blockchain intercommunication is shown in \cref{fig:zkBridgeVision}.

\subsubsection{First contribution: \proto}
Our first contribution is the \acl{dact}\glsunset{dact} (\proto) protocol, which is based on zero knowledge proofs (ZKPs). The \proto protocol is the scheme by which data is transferred in a way that solves the challenges of agnostic anonymous cross-chain messages. When the users transmit a \proto message, no party can tell what the transaction payload is, nor which destination chain the transaction is heading to. Once the transaction has become final on the destination chain, there is no way to link it to the source transaction.

The \proto protocol has two phases: the Deposit phase and the Withdrawal phase. On the Deposit phase, the user creates a \proto message and submits this message to the system that supports the \proto protocol. And on the Withdrawal phase, the user creates the ZKP and performs the settlement of the \proto message through the system that supports the \proto protocol.

\subsubsection{Second contribution: \infra}
\infra is the decentralized infrastructure that supports \proto messages. As \infra works at the application layer of the blockchains, the proposed system is blockchain agnostic. In addition, any dApp can use \infra which makes the system dApp agnostic. Finally, the users can transfer any type of message, with value attached or not, which makes the system message agnostic.

ZKPs break on-chain traceability. Off-chain traceability is broken with an intermediary blockchain that acts as a multiplexer hosting a smart contract (SC) called Mixer. The Mixer stores a global Merkle tree, the Merkle tree leaves are the events from the supported blockchains. These events come from a single SC on each supported blockchain, we call these SC Routers. The objective of a Router is to be a port for the input and output calls from the dApps on each supported blockchain.

\infra integration with the \proto protocol:
\begin{itemize}
  \item \textbf{\proto Deposit phase.} First, the user submits a transaction to the source blockchain dApp SC. The dApp SC performs a business verification of the transaction and sends the user data to the Router SC. The Router SC adds some \proto parameters to the user data. Finally, the Router SC emits an event which is captured by the oracles in their logs. Afterwards, the oracles submit the events to the Mixer SC where the global Merkle tree is stored. Periodically, the oracles update the latest Merkle tree root into all the Router SCs.
  \item \textbf{\proto Withdraw phase.} First, the user creates locally the ZKP for a \proto message and submits it to the destination blockchain Router SC. On the destination Router SC, the ZKP and its public signals are verified. If they are correct, the Router SC makes a contract call to the destination dApp SC with the byte-code data from the user.
\end{itemize}

\subsubsection{First goal: rational players resistance}
The first goal of \infra is to be resistant to rational players from a game theory perspective. This means that we do not follow a reactive security system which are based on an optimistic time window. Instead, the rational players resistance is directed by a decentralized proactive security system. Even if the majority of the oracles collude or if they are hijacked through cyberattacks, the attackers won't be able to fraud the dApps by injecting forged roots. The rationale is that the dApps signs their own user source blockchain event and, via the ZKP circuit, we can know if the dApp has signed a specific message without revealing the message nor the cipher-text, thus preserving the user anonymity.

Furthermore, the dApps can opt to implement a distributed signature scheme to increase their resilience. Going further, if the dApp distributed signature scheme is optimized from a game theory point of view, then the dApps achieve inherent security. This means that the dApp node signers hold no rational sentiments to align with the oracles to maliciously abuse the network. For this to happen, the dApp nodes must be the entities that have something to loose if the oracles were to attack the network.

In other words, if a cross-chain dApp is an asset value transfer that has liquidity pools with a Total Value Locked (TVL) higher than the oracles stake, then the oracles have a rational reason to act maliciously and attack the network. However, if each signer of the dApp distributed scheme is a liquidity provider, then this attack can almost be nullified as the liquidity providers will not align with the rational oracles in order to steal their own money. 

\subsubsection{Second goal: MEV resistance}
The second goal of \infra is MEV resistance. This is achieved as the users do not reveal information at the source channel or in the message tunnel. They only reveal their intentions on the destination blockchain. With this, we achieve MEV resistance for a cross-chain messaging system. But we are still exposed to the MEV from the single blockchain systems as the users have to reveal their intentions on the destination chain. At this point in the destination chain, we can integrate with other existing methods to avoid being subject to a MEV vectors.

\subsubsection{Third goal: privacy}
\infra delivers privacy and anonymity to the users, by two main factors. First, the on-chain graph analysis is broken through Proofs of Membership via ZKPs. Second, the off-chain graph analysis cannot be reconstructed, as we implement a multiplexer blockchain, where we concentrate all the users transactions into a Mixer SC.

\subsubsection{Fourth goal: censorship resistance}
Finally, the fourth goal of \infra is censorship resistance. This is achieved with the privacy that the ZKPs grant and by our design of the global Merkle tree.

The only thing that the oracles know is the source user address, source dApp SC address, and the source blockchain. The oracles have no view of the destination user address nor the destination blockchain, or the payload data.
The oracles can still opt to silence a specific dApp, achieving dApp censorship resistance is for future research. As for blockchain censorship resistance, we have partial resistance. As the only thing that the oracles know, is that a specific package is departing from the blockchain.

\subsubsection{Proof of concept (PoC) implementation}

To minimize efforts and to reuse battle tested code, we have done a Proof of concept (PoC) development by reverse engineering the audited code of \tornado and by applying IDEN3 ZKP circuits. To meet our anonymous agnostic blockchain intercommunication requirements, we have broken down the \tornado components into simple units, that later on we have interconnected and reiterate, to meet our design and code requirements.

\begin{figure}[tbh]
\centering
  \includegraphics[width=0.75\textwidth]{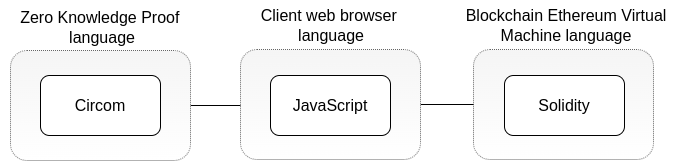}
  \caption{PoC main challenge: Compiler integration to reconstruct the Proof of Membership via the ZKP circuit.}
  \label{fig:languageIntegration}
\end{figure}

The most challenging part of the PoC implementation was the integration of three programming languages, where each language is using different ZKP based libraries. Programs written in these languages must be able to see the same truth, if not, the user funds will be perpetually locked (\cref{fig:languageIntegration}).

\section{Previous Work: Blockchain Intercommunication Protocols}
Blockchain intercommunication protocols, also known as blockchains bridges, listen to transactions from a source chain and settle it on a destination chain. This allows for seamless interaction from one blockchain to another, without requiring users to have native funds on the destination chain. Blockchain bridges unlock liquidity and ease user frictions. For example, one user in the Polygon network that has MATIC (Polygon cryptocurrency) can buy another asset in the Ethereum network without having ETH (Ethereum cryptocurrency). Blockchain bridges can also support the transmission of agnostic payloads, normally called ``messages''. A payload can have value attached to it or an agnostic action such as casting a vote for a DAO.

Blockchains operate on isolated environments to achieve their secure by default design. Blockchain bridges are insecure by definition since they try to connect two isolated environments that weren't designed for interoperability.

The industry created oracles in order to overcome these risks. Oracles are off-chain entities that hold the cryptography keys to perform transaction settlements on the destination chains. There are different approaches that try to ensure the correct behaviour of the oracles.

\subsection{Types of Intercommunication Security Designs}

The main difference among the blockchain bridge architectures relies on their security principles (\cref{tab:bridgeSecurityTypes}). Here, we have two main categories of security principles: proactive and reactive. In cybersecurity, reactive security means that the ``defense team'' must react once the attacker has penetrated its premises. While proactive security focuses on keeping the attacker outside the system boundary.

\begin{table}[tbh]
	\centering
    \caption{Different types of blockchain bridge based security.}
    \label{tab:bridgeSecurityTypes}
    %\vspace{-6pt}
    \begin{tabularx}{.9\textwidth}{ccccc}
        \toprule
        & \bf Proactive security & \bf Reactive security & \bf Security principles\\
        \midrule
        \multirow{1}{*}{\bf Optimistic} & False & True & 1.- Active whistle blowers\\
        & & & 2.- Honest higher hierarchy\\
        & & & 3.- Signed Merkle tree root\\
        \midrule
        \multirow{1}{*}{\bf Reputation} & True & False & Companies with high \\
        & & & reputation are honest\\
        \midrule
        \multirow{1}{*}{\bf Centralized} & True & False & One company is honest \\
        \midrule
        \multirow{1}{*}{\bf ZKP-based} & True & False & Rational players \\
        & & & motivation aren't aligned \\
        \bottomrule
    \end{tabularx}
\end{table}

\subsubsection{Reactive security: optimistic bridge}
An optimistic bridge requires a time window where the transactions are paused until they become final. In the time window, some actors can perform the role of ``whistle-blowers''. If the whistle-blowers see a forged message during the window time, they can flag it as crafted message. Afterwards, the crafted message is externally verified, and if the result is False: then the oracles loose their stake, and the whistle-blower gets the slashed stake from the oracles.

In an optimistic system, there are two ways to validate a transaction. The first option is through a higher hierarchy trusted actor that can be a single off-chain party or a distributed off-chain entity. The second way to verify the transaction validity is through the Merkle Tree Root. With this solution, the oracles obtain the Merkle Tree Root from the source chain, sign the Merkle Tree Root, and input the Merkle Tree Root and the signature on the destination chain. If the oracles submit a forged Merkle Tree Root then any party can read this data from the destination chain and submit the forged Merkle Tree Root with the oracles signature on the source chain. If the Merkle Tree Root does not exist on the source Blockchain Merkle Tree, then the stake of the oracles is slashed.

This unfolds other issues. Users must wait for their transactions to settle which can take days and thus greatly impacts user experience, value transfer, and financial services. Furthermore the higher trusted entity might become erratic and coerce with the oracles, which completely breaks its initial purpose.

Going further, what happens if the crafted message by the oracles has a higher settlement value than the stake that they can loose? In this case, oracles can be Rational Players and attack the network as the spoils of war are high enough to take the risk.

An example of reactive security based on optimistic theory is the Nomad bridge. The Nomad bridge has a security add-on for the dApps to decide which watchers can flag their messages~\cite{nomad}.

\subsubsection{Proactive security: reputation or centralized bridge}
A proactive bridge based on reputation or centralization can settle transactions as fast as they enter into the bridge; there are no time windows. This has clear benefits for the user experience, as the negotiations of asset trading can settle faster, and thus, these bridges are suitable for DeFi projects.

Some projects on the market rely on the honesty of their oracles or on the reputation of their industry. In principle, if a big company is one of the oracles in the network, their reputation is stained if they misbehave. This opens the following questions: Is the system fully decentralized? How do we decide who has enough reputation to become an oracle? Furthermore, what happens if at some point, that entity does not care anymore about their reputation and they abuse the network as a Rationale Player would?

As for the bridges whose proactive security relies on centralization, they inherit the same problems as the reputation security based bridges, plus increased cybersecurity risks. As the attackers only need to orchestrate a cyberattack towards a hand-full of nodes.

An example of a project that is based on a proactive reputation security is Wormhole~\cite{wormhole}. Their security relies on 19 nodes signing the messages and each node is a reputable company on the blockchain industry~\cite{wormholeGuardians}.

\subsubsection{Proactive security: zero knowledge bridges}
Zero Knowledge Proofs (ZKPs) do not make a cross-chain messaging system trustless. However, some cross-chain systems implement ZKPs under the assumptions that the Rational Players will not collude.

In a ZKP cross-chain messaging system, we require an external entity (oracle), to upload ``data'' from the source chain into the destination chain, the ZKP public signals are verified against this data. And if the imported ``data'' is forged, a valid ZKP will be verified as True by the Verifier SC, and the ZKP public signals will match with the forged ``data''. With this, the Rational Players have triumph.

An example of a cross-chain messaging system based on ZKPs is LayerZero~\cite{layerZero}. In LayerZero the oracles upload the source data into the destination chain, this data is required to verify the ZKP public signals. Afterwards, the Relayers create the ZKP with user data from the source chain. Finally, the Relayers submit the proof into the destination chain. The LayerZero Rational Players assumption is that the interests of the oracles and the Relayers are not aligned, thus dApps must set their own Relayer system.

\subsection{Multiplexer Blockchains}
The Multiplexer blockchain concept is applied in the blockchain industry mainly to solve the scalability challenges. This solution is based on connecting Sidechains and the Layer-2 chains to the Layer-1 Ethereum Main Network~\cite{ethl2}. A similar concept is Polkadot through its Relayer Chain and the Parachains~\cite{polkadotWhitepaper}.

Ethereum uses off-chain solutions such as Sidechains and Layer-2 systems to challenge its scalability problems. The collective term of Layer-2 refers to a group of solutions, mainly Rollups and State Channels. The main difference between Layer-2 and Sidechains relies on their security assumptions. On one hand, Sidechains are running in Parallel of Ethereum, but they do not inherit the Ethereum principles of security and data availability. On the other hand, Layer-2 solutions obtain the security principles and data availability from the Ethereum Network.

\begin{figure}[tbh]
\centering
  \includegraphics[width=0.5\textwidth]{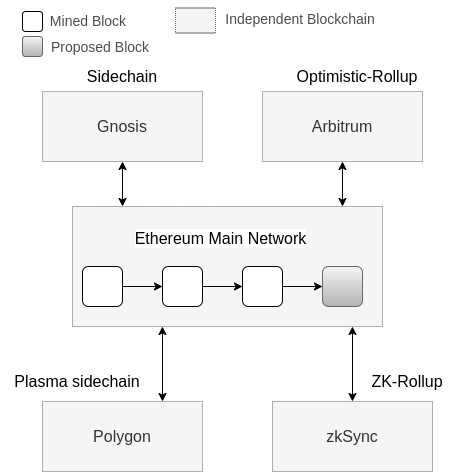}
  \caption{Ethereum as a multiplexer blockchain.}
  \label{fig:ethereum}
\end{figure}

Both, Sidechains and the collective term of Layer-2 solutions are a separate blockchain that is connected to the Ethereum Main network. Therefore the Ethereum Main Network acts as a multiplexer of the other blockchains \cref{fig:ethereum}.

\section{Previous Work: ZKPs Applied to Single-chain Blockchain Privacy}

Projects that provide private payments via \acp{zkp} in a single chain environment have paved the way for anonymous cross-chain communication. These projects range from initial anonymous value transfers with caped usability to full decentralized anonymous payments systems supported at multiple blockchains. In particular, Zerocoin, Zerocash and \tornado have set the foundations of anonymous asset value transfers. However they have different focus. Zerocoin and Zerocash came out as protocol solutions isolated to their own environments. But \tornado is designed at an application level. Public blockchains that didn't have privacy at their protocol level can achieve it at an application level through \tornado SCs. The previous projects focus on private value payments, and not on private agnostic calls. 

\subsection{Zerocoin}
The Zerocoin decentralized privacy paper for Bitcoin came out in 2013~\cite{zerocoin}. Its objective was to provide privacy to its users by breaking the linkability between a transaction settlement and its origin. Its novel contribution was to deliver a decentralized privacy setup, where no centralized issuer was required to preserve the anonymity. The core of Zerocoin is its zero-knowledge implementation to prove that a specific user is the owner of a coin. Zerocoin has some drawbacks~\cite{zerocash}:
\begin{itemize}
  \item High computational costs
  \item No direct payments of native zerocoins among the users
  \item Metadata in the transfer was not private
  \item Coin denomination in the transfer was not private
  \item Fixed coin denominations to obfuscate the transaction graph analysis
\end{itemize}

\subsection{Zerocash}
Zerocash came out as the solution of Bitcoin privacy back in 2014. Its objective is to solve the Zerocoin drawbacks in order to deliver a full fledged anonymous decentralized payment system. To do so, they propose the decentralized anonymous payment (DAP) scheme. Zerocash implements \acp{zksnark}. With this, Zerocash is able to~\cite{zerocash}:
\begin{itemize}
  \item Reduce the transaction size
  \item Reduce the proof verification time
  \item Transfers with arbitrary coin amounts 
  \item Hide the coin value on the transaction
  \item Hide the user's balance sheet
  \item Users ability of direct payments of the coins
\end{itemize}

\subsection{\tornado}
The previous projects of Zerocoin and Zerocash, address the privacy of the Blockchain transactions at a protocol level. On the other hand, \tornado came out in 2019 to provide privacy on the Ethereum chain at an application level through SCs. This means that no changes on the Ethereum Blockchain protocol are required. A user transfers native Ethereum cryptocurrency called ETH into the \tornado SC, and later on the user can submit the \ac{zkp} into the same SC in order to withdraw the fixed denomination value in ETH, each fixed amount is called a coin. With \acp{zkp} and Merkle Trees, \tornado breaks the transaction graph analysis between the deposit and the withdrawal. Furthermore, \tornado expanded its usability to the rest of blockchains compatible with Ethereum, these types of blockchains are named \ac{evm} compatible. As well as with non-EVM compatible Blockchains. The first version of \tornado has the following constraints and principles~\cite{tornado}: 
\begin{itemize}
  \item Fixed amount denomination
  \item User has full custody of its deposit
  \item No Oracles are involved
  \item It is supported in multiple blockchains (EVM and non-EVM)
  \item It works on single blockchain environments
\end{itemize}

\tornado has expanded its support to non native Ethereum cryptocurrency (ERC20). At the time of writing, the team behind \tornado is experimenting with Tornado Nova, which aims to provide transactions with arbitrary amounts and shielded transfers~\cite{nova}.

\section{Background: Zero Knowledge Proofs}

The core concept of \acfp{zkp} is the ability for a prover to prove to a verifier that a specific statement is True without having to reveal any private information. More precisely, it means that a verifier is sure that specific computations were performed correctly without having to execute each step of the process, nor without needing the knowledge of what was executed inside the scheme~\cite{chris}. \Acp{zkp} allow privacy, correctness and validity for a specific data exchange system.

Let's say that Mars is already colonized. Earth and Mars want to strengthen their borders with some new privacy policy. With the new policy, an Earth individual entering the Martian territory must prove to the Martian control border officer, that he has in fact a nationality from Earth. But because of the new privacy policy, the person entering the Martian planet shall not reveal its Earth-ID neither his Earth country to the Martian control border officer. To solve it, we can apply a \ac{zkp} scheme where the validity of the proof grants the affirmation that indeed, the person has an Earth nationality, but without revealing its Earth-ID nor his Earth country. 

Later on, Mars wants to attract more Earthling tourists so it implements a reward mechanism based on the times that an Earthling has entered its borders. In order to comply with the privacy policy, they upgrade the previous \ac{zkp} circuit by adding a public counter and a public alias to each new Earthling. Therefore, every time the Earthling with the same alias enters the Martian territory its reward counter is updated. Later on, this person goes to a third party called martian reward shop and claims its prize according to its public counter, without revealing its Earth Country, nor its Earth-ID. Furthermore, the martian reward shop knows that the proof is valid, and that the reward counter is indeed updated by the martian control border officer. 

\subsection{Different Types of ZKPs}
\begin{figure}[tbh]
\centering
  \includegraphics[width=0.8\textwidth]{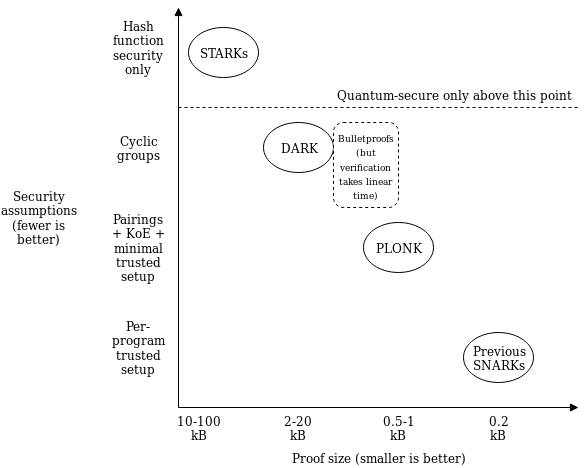}
  \caption{ZKPs comparison: proof size vs security assumptions~\cite{vitalikplonk}.}
  \label{fig:Tradeoffs}
\end{figure}

There are different \ac{zkp} schemes that we can implement. Each of them has its own strengths and constraints. Choosing the right scheme depends on the frequency of transactions to be verified, the proof size, the proof computational time to create it, and the tools that the users have. The smallest proofs are given by \acp{zksnark} (\cref{fig:Tradeoffs}).

Our focus is to be able to prove that a specific action is valid, without the need to have prior interaction with the verifier. This means that a program can play the role of the verifier in the \ac{zkp} scheme. Therefore, an SC deployed on the Blockchain can be called by a dApp to prove if certain data is valid or not, while still preserving the user privacy. This requirement constraints us to implement non-interactive \acp{zkp}. %Zero-Knowledge Proofs (NIZK).

\subsection{Choosing Zk-SNARKs}
We have decided to implement \acp{zksnark}. \Acp{zksnark} have been out in the market since 2012~\cite{chiessasnarks} which has given them more time to have a strong base of developers, community, and libraries. IDEN3 has developed a programming language called Circom, that allows developers to write their own \ac{zkp} arithmetic circuits. Furthermore, Circom has developed JavaScript libraries for the clients to create their proofs locally through their browsers and an export functionality from the verifier key into an SC written in Solidity. This Verifier SC can be implemented in any EVM Blockchain.

\Acp{snark} are succint, they have small proof sizes and sublinear verification time. This means that more complex \ac{snark} proofs won't be exponentially more costly to verify. \Acp{snark} are also non-interactive, the verifier never needs to contact the prover. This means a prover can send its proof in a ``fire and forget'' fashion.

On the other hand, \acp{snark} require a trusted setup in which the \ac{zkp} circuit gets initialized. If the \ac{zkp} circuit is initialized by a single entity, then the users must trust that the entity has destroyed the ``garbage data'' or it could create forged proofs. It's worth noticing that the trusted setup is required a single time during the circuit initialization phase and no further trust is required.

Therefore the trusted setup requirement can be minimized through an open \ac{mpc} where at least one honest entity participating in the Ceremony is required to destroy its ``garbage data''. If there are 100 entities in the open \ac{mpc} ceremony, and 99 of them are malicious, but 1 individual was honest and destroys its ``garbage data''; then the malicious actors won't be able to create evil proofs.

\subsection{SNARKs Arithmetic Circuit}

\begin{figure}[tbh]
\centering
  \includegraphics[width=0.7\textwidth]{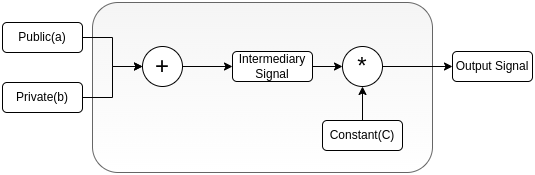}
  \caption{Signals in the \ac{zksnark} circuit.}
  \label{fig:zkpSignals}
\end{figure}

\Acp{snark} only works with numbers as inputs. In order to prove a computational statement (an equation), we must apply a conversion from the equation that we want to prove into an arithmetic circuit. In a high level view (\cref{fig:zkpSignals}), we can see that the arithmetic circuit is formed by multiple wires interconnected through addition and multiplication gates. Through these wires, we have three types of signals~\cite{circomdocs}:
\begin{itemize}
  \item Input signal: The input numbers for our equation
  \item Intermediate signal: The output of each addition and multiplication gate
  \item Output signal: The output signal of the last gate of the circuit
\end{itemize}

The input signals can be private or public but the output signal is always public. Furthermore, the output signal is not mandatory, some circuits do not have output signals. So what is the purpose of a \ac{zkp} circuit without an output signal? Here is where we introduce the concept of assertions and constraints where the proof can only be created if certain value asserts a constraint. For example, the proof can only be created if H[N]===H; where N is the image number, and H is the hash from N.

\subsection{Pedersen Hash with 4-bit Windows}
Hashing is a paramount cryptography function, that we require in order to provide security through an efficient implementation. And hashing is well used within the \ac{zksnark} circuits in order to cover a vast range of use cases.

However, the mathematics behind the hashing functions require advanced techniques, where researchers constantly battle between security and efficiency. The pre-existing Hashing functions did not meet the efficiency standards that we require in order to have effective \acp{zksnark} computation, usability and verification.

Therefore, IDEN3 has implemented the Pedersen hash function due to its efficiency. The Pedersen hash developed by IDEN3 consists in a more efficient version, where they use the Baby-Jubjub elliptic curve and 4-bit windows. With this, they are able to reduce the number of constraints per bit compared to the Zerocash 3-bit window~\cite{pedersenHash}. 

\subsection{Applied zk-SNARKs with Circom}

There are six steps in order to create a seamless user experience of \acp{zksnark} with Circom. The order is as follows: 
\begin{enumerate}
  \item Compile the equation into the Circom language.
  \item Set up the Circuit, so we can create the Verification and Proving Key.
  \item Export the Verification function and Verification key into an EVM SC.
  \item Compute the witness with the signal inputs. The Witness allows us to translate the input signals with its private inputs, into a data structure that allows verification of the ZKP circuit statement.
  \item Generate the proof, this outputs the proof and the public signals.
  \item Validate the proof, either locally or in the Blockchain through the Verifier SC.
\end{enumerate}

In Circom, one of the stages where we have to apply MPC is on the Setup of the circuit. During the Setup phase, we apply the trusted setup of the Groth16 \ac{zksnark} scheme~\cite{groth16}. The MPC ceremony has two main phases, and it can be done through the JavaScript (JS) Application Protocol Interfaces (APIs) that Circom has developed.~\cite{circomdocs}:
\begin{enumerate}
  \item Circuit independent: Powers of Tau.
  \item Circuit specific: Phase 2.
\end{enumerate}

It is important that both parties in the \ac{zkp} scheme are coordinated, and in order to do so, both parties shall compile the same circuit. In order for the Verifier to know that the Prover won't be malicious, the Verifier sends his Circuit Proving Key so that the Prover can produce the Proof with the certainty that it matches the Verifier Circuit. Afterwards, the Verifier obtains the Prover proof and verifies it through his Verify Key. The output will be True if the proof is valid and False otherwise.

\begin{figure}[tbh]
\centering
  \includegraphics[width=0.999\textwidth]{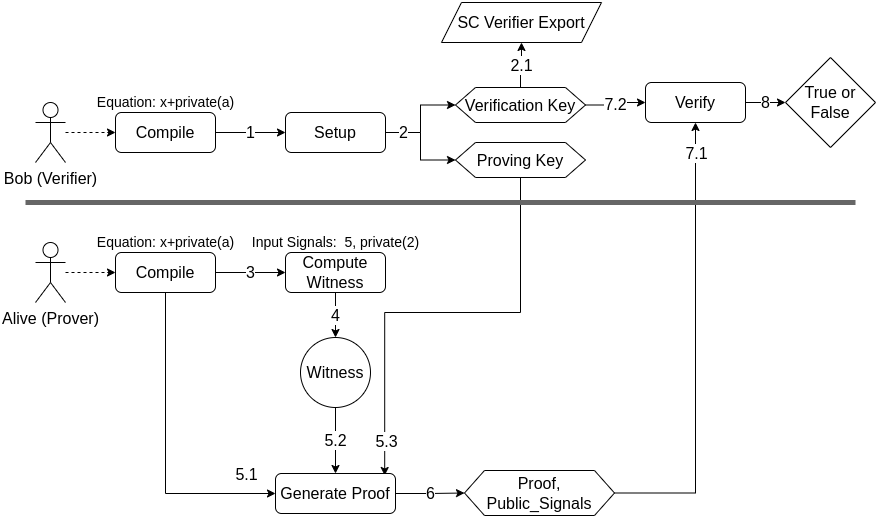}
  \caption{Applied zk-SNARKs with Circom.}
  \label{fig:zkpApplied}
\end{figure}

In \cref{fig:zkpApplied} Alice and Bob agree on a specific arithmetic scheme that they want to apply (A+B), without revealing specific information (B). In this case, Alice (prover) must convince Bob (verifier) that her statement is true. Everything above the middle line is what Bob knows, and everything under the middle line is what Alice knows.

The steps of \cref{fig:zkpApplied} are as follows:
\begin{enumerate}
  \item Alice and Bob compile the exact same equation into the circuit.
  \item Bob makes the Setup ceremony trough a MPC, and obtains the Verification Key and the Proving (they only work with this circuit). Bob exports the Verification Key and the Verification function into an EVM compatible SC written in Solidity.
  \item Alice computes the Witness of her inputs the signals of x=5 AND private(A=2).
  \item Alice has her Witness.
  \item Bob sends his Proving key to Alice. Alice generates the proof with the: Circuit compiled output, her witness, and Bob's Proving Key.
  \item Alice send the proof to Bob. The output proof from Alice contains two files: the proof and the public signals.
  \item Bob verifies Alice proof with his Verify function that calls his Verification Key.
  \item If the output is valid, it means that Bob knows that Alice used the input signal of x=5, but he has no clue about the private A signal. 
  
  If the output is invalid, it means that either Alice used a different circuit, or that Alice changed something inside the proof file or in the public signal file so Bob knows that Alice is an evil actor.
\end{enumerate}

The design of private signals must be in such a way that the user cannot easily deduct them. In \cref{fig:zkpApplied}, the private signal can easily be obtained by finding the variable in the equation.

\section{The DACT Protocol}

Our first contribution, as a global solution for the anonymity and security challenges of agnostic Blockchain intercommunication, is the \acf{dact} protocol. The users can send data from one chain to another using \proto messages which are the means of data transfer. \proto messages allow to maintain privacy in a cross-chain scenario.

The \proto protocol implements a verification system based on Proof of Membership via Zero Knowledge Proofs (ZKPs). Via the ZKP, the user proves that it knows a specific path to a leaf from the Merkle Tree, without revealing which Merkle Tree Leaf it is. Furthermore, the ZKP also verifies that the user Merkle Tree Leaf has been signed by the dApp owner (single or distributed signature) without revealing the cipher-text nor the message.

\subsection{\proto Data Structures}
The \proto protocol has several data structures. For simplicity, we have structured them accordingly to the entity that creates the data structures.

\subsubsection{User custody data structures}
The user has custody of its transaction as the user can revert its funds independent from the Oracles and the dApp. The user is the one that creates the data structures that allow to reconstruct its commitment and the Merkle Tree Leaf. With this, the user can prove that it knows a path to that Merkle Tree Leaf.

User Custody data structures created locally:
\begin{enumerate}
  \item \textbf{Secret}: BigInteger from random 31 bytes. To raise up the security bar.
  \item \textbf{Nullifier}: BigInteger from random 31 bytes. To raise up the security bar.
  \item \textbf{Salt}: BigInteger from random 31 bytes. To obfuscate the Users intentions data.
  \item \textbf{Nullifier Hash}: BigInteger, Pedersen-Hash from the Nullifier. To avoid double spend.
  \item \textbf{User obfuscated data}: Keccak256-Hash of the agnostic payload in Bytes 32, the destination chain ID and the Salt. To avoid the user from changing its intentions and to obfuscate the user intentions. Pre-input from the Merkle Tree Leaf.
  \item \textbf{User Commitment}: BigInteger of the Pedersen-Hash from the Secret and the Nullifier. To obfuscate the Secret and the Nullifier. Input to the Merkle Tree Leaf.
  \item \textbf{ZKP}: SNARK proof from the \proto circuit. To deliver a Proof of Membership.
  \item \textbf{User Withdrawal call parameters}: ZK proof and the Payload in Bytes32. To execute the settlement of the initial transaction.
\end{enumerate}

\subsubsection{\proto SC data structures}
For security, the \proto protocol SCs shall create the structures that we do not want the user to submit by himself to ensure that the user cannot game the system.

\proto SC data structures:
\begin{enumerate}
  \item \textbf{Destination dApp SC address}: EVM hexadecimal address.
  \item \textbf{Version}: Integer from 1 to 1000. To grant backward compatibility.
  \item \textbf{Source Chain ID}: Integer from 1001 to 10000. To avoid double spend on the Revert transaction. Input to the Merkle Tree Leaf.
  \item \textbf{Destination Chain ID}: Integer from 1001 to 10000. To avoid double spend on the transaction settlement.
  \item \textbf{dApp Global Hash}: Keccak256-Hash of all the dApp SC addresses in all the blockchains. Input to the Merkle Tree Leaf.
  \item \textbf{Trustless public commitment}: BigInteger of 73 bits, from the Keccak256-Hash of the dApp Global Hash, the version, and the User obfuscated data. The objective is to concentrate public multiple data into one single public ZKP public signal. This is an input to the Merkle Tree Leaf.
  \item \textbf{Merkle Tree Leaf}: Bytes32, Addition of the User commitment, the trustless public commitment, and the source Chain ID.
  \item \textbf{Merkle Tree Root}: Bytes32 string.
  \item \textbf{Merkle Tree Path Elements}: Array of bytes 32. The length is the Merkle Tree Height.
  \item \textbf{Merkle Tree Path Indices}: Array of 1 and 0. The length is the Merkle Tree Height.
\end{enumerate}

\subsubsection{dApp SC data structures}
In order to provide a decentralized proactive security. The dApp must construct the leaf signature structure. With this, a cyberattack or a Rational Player must obtain control from the Oracles threshold nodes, and of the dApp node.

dApp SC data structures:
\begin{enumerate}
  \item \textbf{dApp Signature}: ECDSA signature of the dApp from the user leaf. To avoid valid forged ZKP settlements.
  \item \textbf{dApp Public Key}: ECDSA Public Key. To avoid valid forged ZKP settlements.
\end{enumerate}

The signature structure is sent to the user, either through the Mixer SC, or through another communication channel. In order to reconstruct the ZKP, the user requires the dApp signature.

\subsection{\proto Proof of Membership}
The proof of membership from a specific Member (data) is delivered via the correct formation of the ZKP and the assertions of the ZKP public signals. In other words, the Merkle Tree Leaf is the Member object and the Merkle Tree is the Membership group. Therefore, the user is able to prove that the Member (Leaf) held by the user belongs to the Membership group (Merkle Tree) without revealing which Member (Leaf) it is. 

\subsection{\proto ZKP Circuits}
\begin{figure}[H]
\centering
  \includegraphics[width=0.75\textwidth]{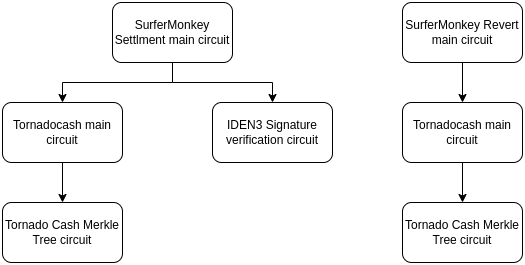}
  \caption{\infra ZKP circuit hierarchy: Settlement circuit (left), Revert circuit (right).}
  \label{fig:zkpCircuitHierarchy}
\end{figure}

The \proto protocol gives to the users full custody of their funds and avoid malicious behaviour by proof replication on double spend attacks (withdraw and then revert). The \proto protocol has two independent main circuits (\cref{fig:zkpCircuitHierarchy}): the Settlement circuit and the Revert circuit.

The settlement circuit is a triple nested circuit. From \tornado, we obtain two implementations and concepts: the circuit shall prove that the user knows how to make a leaf, and that it knows the path to reconstruct the Merkle Tree Root, without revealing that specific Merkle Tree leaf. The third circuit is from IDEN3, we use the signature verification circuit.

The Revert circuit is also based on the \tornado circuits. However, here the user breaks the anonymity by revealing the commitment and the source chain ID.

Our first circuit deviation from \tornado is the signature verification circuit developed by IDEN3. With this we raise up the security bar against cyberattacks and Rational Players. Therefore, even if the Oracles collude or are hijacked and insert forged roots into the \proto SCs, the attackers and Rational Players won't be able to settle transactions. By adapting this circuit to our system, we have certainty that the dApp has indeed accepted and signed a specific user leaf, but without revealing the leaf nor the signature.

Our second circuit deviation from \tornado is the modification of the Merkle Tree Leaf. In our case, the Merkle Tree Leaf is the addition of the Source Chain ID, the Trustless Public commitment, and the commitment.

\subsubsection{\proto ZKP Settlement circuit}
The Settlement ZKP circuit has the input signals from \cref{tab:signals-vs-tornadocash}. We differentiate the signals that were placed by \tornado from the signals added for \proto. \proto ZKP circuit signals:

\begin{table}[H]
	\centering
    \caption{Comparison of public and private signals between \tornado and \infra.}
    \label{tab:signals-vs-tornadocash}
    %\vspace{-6pt}
    \begin{tabularx}{.8\textwidth}{ccc}
        \toprule
        & \bf \tornado & \bf \infra \\
        \midrule
        \multirow{2}{*}{\bf Public signals} & Nullifier hash & Trustless public commitment \\
                                & Merkle tree root & dApp public key\\
        \midrule
        \multirow{4}{*}{\bf Private signals} & Nullifier & Source chain id \\
                                & Secret &  Merkle tree leaf signature  \\
                                & Merkle tree path elements & \\
                                & Merkle tree path indices & \\
        \bottomrule
    \end{tabularx}
\end{table}

\subsubsection{\proto ZKP Revert circuit}
The Revert function grants the user the ability to Revert a transaction without interacting with the dApp nor the Oracles. However, it comes at the cost of breaking the anonymity of the user but not its privacy (the user intentions remain obfuscated).

Via the Revert Circuit the users have full custody of their transaction. This ZKP allows the user to revert a function before it has settled on the destination chain. For example, a user deposits 100 ETH in Blockchain A and decides, for any given reason, to revert the transaction and obtain the funds back. The Revert ZKP circuit is similar to the Settlement circuit though it does not implement the dApp signature verification and it has as public signals: the commitment and the source chain.

It only works on the source chain where the commitment was originated. To avoid the Oracles from reverting arbitrary transactions on the source chain and to prevent a user or Oracle to perform a double spent attack, we require the users to create a ZKP with another circuit called ``Revert''. Same as the Settlement-ZKP, the Revert-ZKP can only be created by the user that has custody of the commitment structures.

\subsection{\proto Security}
We have placed several security constraints and asserts inside the \proto scheme. These security principles range from user misbehaviour, user gaming the system, increase the resilience against cyberattacks, to Oracle collusion (Game Theory).

The transfer starts with the user creating its commitment and its user obfuscated data. These data structures form a partial Merkle Tree Leaf. Therefore, the user cannot change any of the initial parameters, as this will output a different Merkle Tree Leaf, and thus the user won't be able to reconstruct the Merkle Tree Root.

\subsubsection{Proactive security}
The Proactive security \proto principle is given by the dApp signing the Merkle Tree leaf, if the dApp has not signed the Leaf, then the users cannot create the ZKP. This provides resistance against Rational Players (Game Theory) and raises the security bar against cyberattacks.

From a Game Theory point of view. If the Oracles collude and they create forged Merkle Tree Roots and inject those roots into the destination Blockchain SC, then the Oracles can create valid ZKPs and perform forged settlements. For example steal funds from the dApps. Therefore, the dApp signature is required for the creation of the ZKP, the signature and the Merkle Tree Leaf remain as private signals. One of the public signals from the ZKP is the dApp public key. This public signal is mapped on the destination Blockchain \proto SC to obtain the dApp Global Hash. This dApp Global Hash must match the dApp Global Hash that recreates the Trustless Public commitment, which is a Public Signal on the ZKP.

From a Cybersecurity principle, by decentralizing the signature from Oracles to Oracles and dApp, the attackers must gain control not only from the Oracles consensus threshold machines, but also the threshold of nodes inside the dApp signature. 

\subsubsection{Resilience to double spends through the Revert function}
Only the user that has the \proto commitment can recreate the Revert-ZKP, and thus pass the parameters to the Revert function. The Revert function can solely be called on the correct source chain ID, this means that the Revert-ZKP public signals match the \proto commitment and source chain ID from the source Blockchain. The revert function implements a time window for the action to be executed. During this time window the dApp can halt the \proto Revert function. If the time window naturally expires, the Revert function is executed.

During the time window, the dApp can analyze the destination chain and check that the user has indeed flagged the \proto Nullifier-Hash as Reverted. If the user hasn't flagged the Nullifier-Hash as reverted on the destination chain, or the Nullifier-Hash is flagged as spent. Then the dApp can halt the Revert function.

Within the Reverse Function, the motivation of implementing a reactive security system based on optimistic theory, where if nothing happens the transaction is executed, has three foundations. First, it motivates the dApps to implement more nodes as they want to have a higher degree of liveliness in the network. Second, it persuades dApps to be constantly vigilant. Third, if by any reason the dApp nodes are annihilated, the user can still Revert the transaction.

The opposite security is that the dApp would have to green flag honest \proto reverse transactions. If the dApp does not green flag a transaction, then the transaction gets automatic reverted. However, this antithesis security would lead to a lazy dApp behaviour.

The Revert Function can be called by anyone. To avoid spams, the \proto Revert Function has a financial expense and it implements a ``cool down period'' during which the user cannot call the function.

A user can lose its funds because the dApp owner node signer is offline or because the dApp has become evil and doesn't want to return the funds to the user. To avoid loses, it is of high importance that the dApp implements a Distributed Signature Scheme where the signing nodes are the entities who desire two things: User success, and that won't have sentiment to align with the Oracles or with a another party, in order to abuse their dApp or its users.

\subsubsection{dApp resilience rules within the revert function}
To secure dApps even in case where their nodes had been totally annihilated, the dApps can implement \proto Revert frequency rules and \proto Revert value Thresholds. With this, dApps can have pre-written rules, of how many Revert transactions they can have per hour, per day or per week. And in the case of value transfer, they can place value thresholds per transaction or per time flow. The revert value thresholds shall be equal or less than the Maximal possible single deposit.

All in all, the resilience rules grant to the dApp administrator and users time to perform salvage operations such as the Liquidity Providers withdraw their funds from the compromised dApp.

\subsubsection{Commitment}
Security constraints inside the user commitment. The commitment is the Pedersen-Hash from the Secret and the Nullifier.

Commitment security principles:
\begin{enumerate}
  \item \textbf{Secret}: Increase the security of brute-force against a Merkle Tree Leaf.
  \item \textbf{Nullifier}: Increase the security of brute-force against a Merkle Tree Leaf. And is also the image from the nullifier hash.
\end{enumerate}

\subsubsection{dApp global hash}
The goal of the dApp global hash is to maintain privacy and to provide security for the dApps. We cannot trust the users to behave correctly and place the correct destination dApp SC address alias. Therefore, the \proto SC is the one that inputs the dApp global Hash (dApp global alias) into the Merkle Tree Leaf.

From a security point of view, one user could deploy a dApp and call another dApp SC address to steal its funds. To avoid this, we are constraining the users to only interact within the same dApp. It is the dApp responsibility to verify the transaction validity on their source SC dApp. For example, the right amount of funds have been deposit.

From a privacy point of view, if the source Blockchain SC was placing the destination dApp SC address into the Merkle Tree leaf, then we will be breaking the user privacy as we would be revealing information about the Source and Destination Blockchain. Therefore, to maintain privacy we introduce a global unique Hash to each dApp.

The creation of the dApp Global Hash happens when the dApp owner submits via the dApp SC a Registry call to the \proto SC. The parameters of this call is an array of all the dApp SC addresses in all the Blockchains except the one where it is calling from. Then, the \proto SC makes the Keccak256-Hash from the SC addresses array and the SC address from where the dApp SC that is performing the call.

\subsubsection{Source chain ID}
The motivation behind implementing the source chain ID into the Merkle Tree Leaf is to avoid reverting a transaction from a different source Blockchain.

\subsubsection{Merkle tree leaf}
\begin{figure}[tb]
\centering
  \includegraphics[width=0.999\textwidth]{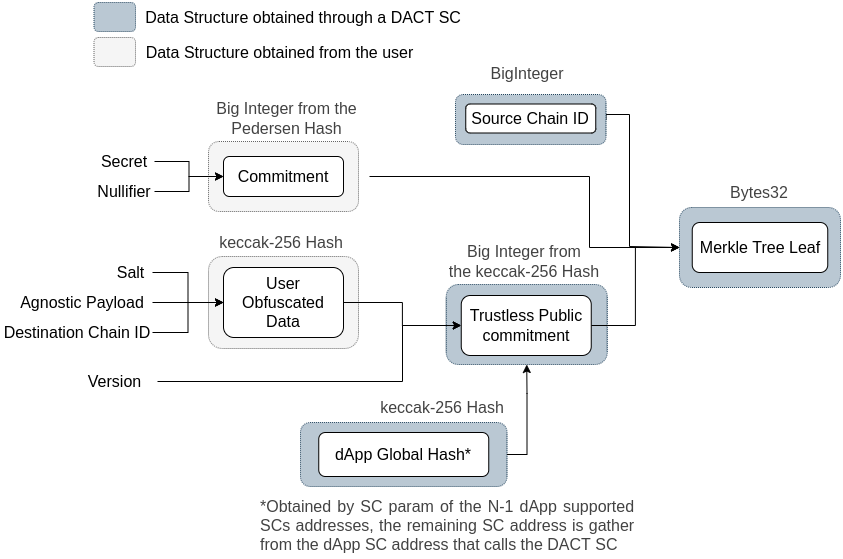}
  \caption{\proto: Merkle tree leaf creation.}
  \label{fig:MerkleLeafCreation}
\end{figure}

The \proto Merkle Tree Leaf is created via two main principles: data that the User must input and data that we cannot trust the user to input. The data that we cannot trust the user to input is placed by the \proto SCs (\cref{fig:MerkleLeafCreation}).

Merkle Tree Leaf security:
\begin{enumerate}
  \item \textbf{Commitment}: To raise the security bar against Brute Force Attacks.
  \item \textbf{Trustless public commitment}: To constrain the user to interact with a single destination chain, within a specific dApp, and with specific intentions.
  \item \textbf{Source chain ID}: To avoid revert from a different source Blockchain.
\end{enumerate}

\subsection{\proto Creation and Settlement}
The \proto has two primary phases. The first phase is named Deposit and the second phase is called Withdraw.

\textbf{\proto Deposit phase}
\begin{enumerate}
  \item The user creates the random numbers: secret, nullifier and salt.
  \item The user creates the commitment which is the Pedersen-Hash from the secret and the nullifier. Then, the user creates the obfuscated data parameters which is the Keccak256-Hash of the Nullifier-Hash, the agnostic payload, and the destination chain ID.
  \item The user submits its commitment, the obfuscated data, and the version into the source Blockchain dApp SC. The source Blockchain dApp SC passes these data to the \proto SC.
  \item The \proto SC creates the Merkle Tree Leaf by making the addition of the previous data.
  \item The dApp Owner checks which leaves are from its own dApp, and through an off-chain process it signs its own leaves and submits the signature to the \proto SC.
  \item Periodically the supported Blockchains get the updated with the latest Merkle Tree Root.
\end{enumerate}

\textbf{\proto Withdrawal phase}
\begin{enumerate}
  \item The user obtains the Merkle Tree Leaf signature from its transaction and the user creates the  ZKP.
  \item The user sends the ZKP to the destination Blockchain \proto SC.
  \item The destination \proto SC checks the validity of the ZKP public signals and that the ZKP is verified as true. Finally, the \proto SC calls the destination dApp SC address. At this point, the transaction is settled.
\end{enumerate}

\subsection{\proto Unlinked Finality and Atomicity}
The \proto transaction finality is given through a virtual network of the supported Blockchains. Here, we can identify which Nullifier Hash has been settled.

Atomicity means that either all the steps from a transaction are executed correctly or that no changes occur. We achieve this through the \proto Revert function, as if the \proto transaction failed and the Unlinked Finality was not achieved, the user can Revert the \proto transaction.

We are interconnecting isolated environments that have no native intercommunication path among them. Once the transaction has been settled on the destination chain, the source chain has no visibility on the state of the destination chain. If we replicate the settlement transaction state from the destination chain on the source chain. We will be breaking the user privacy as the source Blockchain and destination Blockchain would be linked.

\section{The \infra Infrastructure}

\begin{figure}[tb]
\centering
  \includegraphics[width=\textwidth]{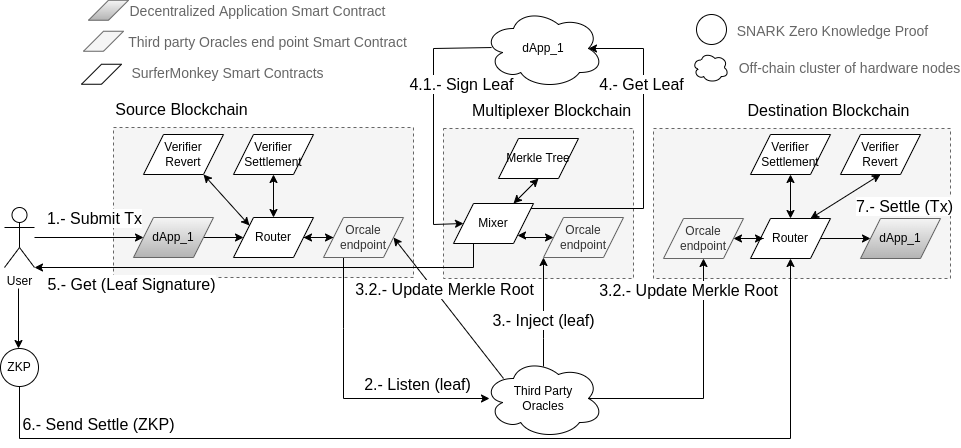}
  \caption{\infra simplified flow.}
  \label{fig:SurferMonkeyTopView}
\end{figure}

Our second contribution is \infra, the decentralized infrastructure that supports the \proto messages (\cref{fig:SurferMonkeyTopView}). \infra is formed by placing Router SCs on each supported Blockchain that can receive \proto messages. During the \proto Deposit phase, the Router SC obtains the Source Chain ID and the dApp Global Hash to create the \proto Trustless Public Commitment. During the \proto Withdraw phase, the Router SC calls the Verifier SC to assert the validity of the ZKP. Furthermore, we also use a multiplexer Blockchain where we concentrate and inject all the messages into a Mixer SC. The Mixer SC has an interface with the Merkle Tree SC that stores the global Merkle Tree where each input transaction becomes a Merkle Tree Leaf.

As for the Rational Players connection, the third party Oracles are connected to \infra via their own SC endpoints. The dApp distributed nodes network interacts with the system via the Mixer SC. As an initial example, the dApp stores the message signature on the Mixer SC but this could be done trough an off-chain process while preserving the user privacy and anonymity.

Periodically, the third party Oracles get the latest Merkle Tree Root from the Mixer SC and update this Merkle Tree Root into the Router SCs. The Proof of Membership from the users that is created via the ZKP circuit is verified against the Merkle Tree Root. The ZKP circuit grants to the system certainty that the user has created the Proof of Membership following specific rules.

The \infra system works at the application layer from the Blockchains (Blockchain agnostic), which means that any dApp can use \infra (dApp agnostic), and transact any type of message (message agnostic).

We break the on-chain and off-chain graph analysis, in other words the inputs and outputs cannot be linked among them. With this, no entity can tell from a transaction what is the destination Blockchain nor its payload. Once the transaction has settled on the destination chain, we cannot reconstruct the path to link it with the initial transaction.

\subsection{System Requirements and Design Principles}
From the current Blockchain industry needs we obtained the System Requirements. And we follow the design principles that will give to the system a higher security level.
System requirements in order of importance:
\begin{enumerate}
  \item \textbf{Decentralization:} No one party can individually kill the system.
  \item \textbf{End to end anonymity:} Input transactions cannot be linked to a destination. And Output transactions cannot be linked to an Input transaction.
  \item \textbf{Blockchain graph analysis anonymity:} The Blockchain ledger and mempool reveal no information about the transaction path analysis.
  \item \textbf{Off-chain graph analysis anonymity:} We cannot know where a specific package flow is going to or coming from.
  \item \textbf{Blockchain, dApp, and message agnostic:} It works on any type of Blockchains, dApp or message type, either with value attached or not.
  \item \textbf{User transaction custody:} The user has custody of its funds, no party can steal or lock the user funds.
  \item \textbf{Revert functionality:} The user can cancel a transaction before it gets settled on the destination Chain.
  \item \textbf{Transaction finality and Atomicity:} Each \proto message transaction shall have an end, and it shall be done in such a way that either all the computational settlement steps are executed, or none of them are.
  \item \textbf{Rational Players resistance:} The actors from the system cannot collude in order to obtain a higher reward.
  \item \textbf{MEV resistance:} The users are shielded from MEV Rational Players.
  \item \textbf{Censorship resistance:} The Oracles or Blockchain Miners cannot silence a specific user.
  \item \textbf{Simple blockchain integration:} To support as many Blockchains as we can, the integration shall be simple. For example, one SC as port.
  \item \textbf{Backward compatibility:} Previous versions of the system shall keep working after the system is updated.
\end{enumerate}

As for the design principles that we placed during the analysis and the creation of the architecture, the most important one for us was Secure by default.
Design principles in order of importance:
\begin{enumerate}
  \item \textbf{Secure by default:} We have implemented the must secure solution that we have analyzed. The security is not an optional principle, it is assured without the users knowing that it is there. Our security principles are added to treat the root cause of the problems, not its symptoms \cite{secureByDefault}.
  
  First, the system works securely as long as the dApps use a Distributed Signature Scheme and the entities that run those nodes hold no rational motivations to align with evil Actors. If the dApp nodes are offline, no forged settlement transactions can occurred. Furthermore, even if the dApp and Oracles are offline, the user can still Revert the transaction and obtain its funds back.
  
  Second, the Merkle Tree Leaf creation follows a holistic approach. We take into account the user intentions and skills, the technology, the process and governance. Therefore, any parameter that creates the Merkle Tree Leaf that can game the system is input by a \proto protocol SC.
  
  Third, the dApp Global Hash cannot be hijacked as one parameter to create it is the SC address that calls the Router SC and this function can only called by the dApp owner.
  \item \textbf{Data integrity:} The Blockchain provides immutability to the stored data. Therefore, it is of primary importance that the all the used languages can recreate the truth that is stored in the Blockchain.
  \item \textbf{Data availability:} The Merkle Tree data is saved on the Blockchain.
  \item \textbf{Composability:} The system follows a modular architecture, so each module can be interconnected following a different pattern. It allows for a simpler upgradability.
  
  An example of simple upgradability, is done through the \proto protocol structure of the User Obfuscated Data. Here, we can input more information in the future, and it does not change anything in the architecture.
  
\end{enumerate}

\subsection{System Architecture}

\begin{figure}[tb]
\centering
  \includegraphics[width=0.999\textwidth]{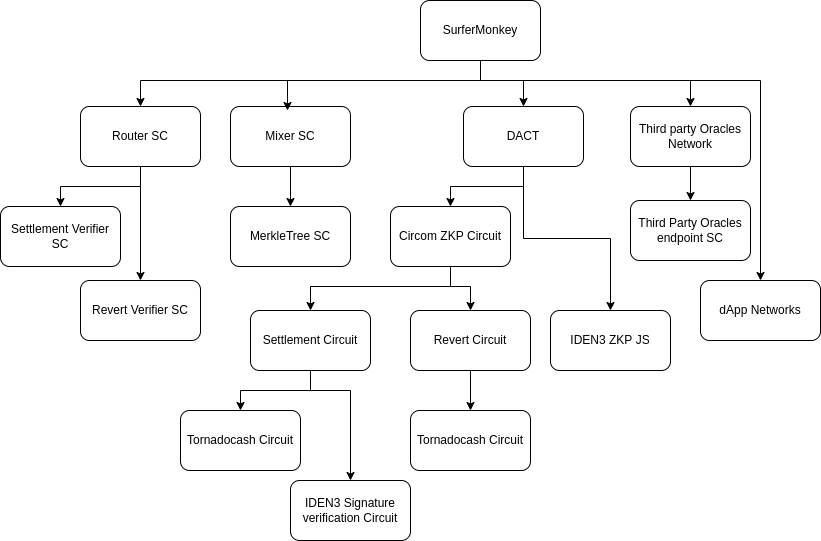}
  \caption{\infra subsystems.}
  \label{fig:surferMonkeyOverview}
\end{figure}

In \cref{fig:surferMonkeyOverview} we have the Surfer Monkey System Architecture which is composed of various subsystems following a hierarchy tree. The components range from JavaScript (JS) libraries, Smart Contracts (SCs), and ZKP circuits. There are also off-chain entities such as Oracle networks and dApp networks. 

\subsubsection{Actors}
The Actors, defined as off-chain entities from our system, are the Users, dApps, and the Oracle Network.
\begin{enumerate}
  \item \textbf{Users:} End users that have the need to perform a cross-chain communication while remaining private.
  \item \textbf{dApps:} Applications that offer a specific need to the users.
  \item \textbf{Oracles:} A network of nodes that gather all the Merkle Tree Leaves and inject them into the Global Merkle Tree. Furthermore, they periodically update the Merkle Tree Root into the supported Blockchains.
\end{enumerate}

We have modelled each of these actors as Rational Selfish Players, that will always look for their best own interests, either for the long run or in the short term. We highlight, that we have taken into account that the different actors can try to maliciously cooperate with each other in order to obtain a higher reward.

Actors by Game Theory definition:
\begin{enumerate}
  \item \textbf{Users:} They want to perform double spend attacks by reconstructing forged withdrawals and by performing Withdraw-Revert attacks.
  \item \textbf{dApps:} They can block a settlement or a Revert function in order to obtain a higher economical value. They also want to reconstruct forged proofs.
  \item \textbf{Oracles:} To obtain ownership of the system, and therefore perform unlimited forged settlements, this entity will try to inject forged Merkle Tree Roots.
\end{enumerate}

Therefore, we have taken into consideration the Rational Players selfish motivations, and we have set several security constraints into the system.
\begin{enumerate}
  \item \textbf{Users:} They have custody of their ZKPs and therefore no one else can reconstruct them. The ZKP circuit avoids double spending and the creation of forged ZKPs. Furthermore, we avoid data replication by using two ZKPs: one for the Settlement transaction, and another one for the Revert Transaction.
  \item \textbf{dApps:} First, dApp nodes sign each User Leaf from their own dApp. Second, a Distributed Signature Scheme is implemented where each of the share holder node has rational intentions for the success of the dApp. As the dApp signers will not steal from themselves, we obtain ensure that the dApp signers will not cooperate with the Oracles in order to form forged settlements. For example, each signer is a Liquidity Provider from a Value transfer dApp.
  \item \textbf{Oracles:} Oracles are not able to reconstruct a ZKP without the dApp Distributed Signature on a Merkle Tree Leaf since dApp nodes do not have rational motivations to maliciously cooperate with the Oracles. Therefore, there is no point in reconstructing a forged Merkle Tree Root. 
\end{enumerate}

\subsubsection{Mixer SC}
The Mixer SC lives on the multiplexer Blockchain. Its main objective is to provide an interface with the Merkle Tree SC.

The Mixer SC facilitates the interaction with the Merkle Tree SC. The Mixer SC provides a link to inject the Merkle Tree Leaves and to obtain the latest Merkle Tree Root. Only the Oracles can call the function to inject a Merkle Tree Leaf. In order to avoid double submission for the same user commitment, the Mixer SC has a mapping function that stores all the user commitments.

\subsubsection{Merkle tree SC}
The Merkle Tree SC objective is to form a Merkle Tree. This SC is cloned from \tornado Merkle Tree SC. It supports up to 32 levels of depth. It has an interface to call the MiMCSponge SC, which implements the block cipher and hash function of Minimal Multiplicative Complexity (MiMC)~\cite{mimc}. The MiMC sponge constructs the MiMCHash. This hash is obtained from two leaves from the Merkle Tree. The objective of the MiMC function is to be efficient in a SNARK circuit environment. 

\subsubsection{Router SC}
Each supported Blockchain has a Router SC that stores a list of all the dApp Global Hashes and that works as a port of input and output transactions. On the \proto Deposit phase, the Router SC obtains the dApp Global Hash and the Source Chain ID to create the \proto Trustless Public Commitment. On the \proto Withdraw phase, the Router SC checks the assertion of the public signals from the ZKP and the ZKP validity. To check the ZKP validity, the Router SC implements an interface to call the Verifier SC for both the Settlement ZKP and the Revert ZKP.

Furthermore, the Router SC also has a function where the dApp List is updated with the dApp Global Hash against the dApp SC address on the same Blockchain as the Router SC. This function is called by the dApp Node Owner via the dApp SC.

\proto Deposit phase - Router SC checks:
\begin{enumerate}
  \item Commitment hasn't been submitted (for User Experience, the one that is for security is the Commitment mapping at the Mixer SC).
  \item dApp Global Hash exists.
\end{enumerate}

\proto Withdraw phase - Router SC checks:
\begin{enumerate}
  \item Withdrawal is not double spending.
  \item Merkle Tree Root is known.
  \item Correct Chain ID, either for destination settlement or source revert.
  \item Payload hash recreates the ZKP public signal of the Trustless Public Commitment. In the \proto Settlement ZKP, the Destination Chain ID is public through the creation of the Trustless Public Commitment. Only on the \proto Revert ZKP, the Source Chain ID is public.
  \item Proof validity.
  \item Obtain dApp SC address from a mapping from the dApp Global Hash. The Trustless Public Commitment is recreated using the dApp Global Hash. 
\end{enumerate}

Periodically, the Oracles update the global Merkle Tree root into the supported Routers SCs with the latest Merkle Tree Root from the Mixer SC. The Router SC only stores the latest two Merkle Tree Roots.

The update period of the latest Merkle Tree from the Mixer SC into each Router SC can range from a few minutes to several minutes. The main constraints on updating the Mixer Merkle Tree Root into all the Router SCs are the computational expenses and the transaction settlement time to become final on each supported blockchain. For example, there is no reason to update every 5 minutes the Merkle Tree Root in Blockchain A if the transaction finality in Blockchain A is granted every 30 minutes. 

\subsubsection{Verifier SC}
The Verifier SC is obtained from an export functionality from the Circom language. This SC has the verify function with the verify key from a specific \ac{zksnark} circuit. 

\subsubsection{dApp SC}
This SC falls outside the boundary of the \infra system. The dApp places its business logic on this SC. On the \proto Deposit phase, the dApp passes the user commitments to the Router SC. On the \proto Withdraw phase, the dApp SC gets called by the Router SC. Finally, the dApp SC must implement a Revert functionality in order for the users to recover their funds.

dApp SC requirements:
\begin{enumerate}
  \item \proto Deposit phase: Interface to call the Router SC.
  \item \proto Withdraw phase: Grant permission to be called by the Router SC.
  \item Implement a revert function. For example, in case of value transfer the users can revert the funds back to the initial sender account.
\end{enumerate}

\subsubsection{\proto}
The \proto is the message transfer protocol that \infra implements. Through the \proto protocol we have solved the challenges of anonymous agnostic Blockchain intercommunication. The \proto circuit is based on a circuit developed by \tornado.

\subsection{Settlement Transaction Flow}
\begin{figure}[H]
\centering
  \includegraphics[width=0.99\textwidth]{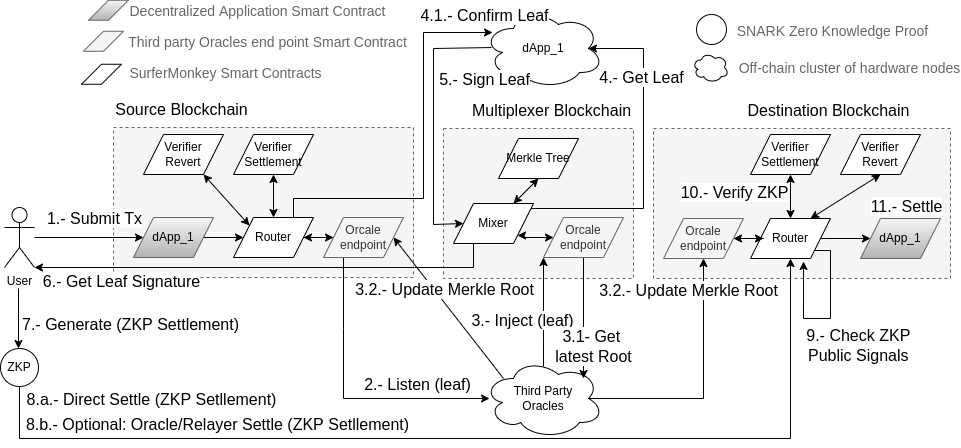}
  \caption{\infra: \proto Settlement transaction.}
  \label{fig:SurferMonkeyDiagram}
\end{figure}

The end to end flow of a \proto Settlement transaction through the \infra architecture (\cref{fig:SurferMonkeyDiagram}), is performed in two main phases: \proto Deposit and \proto Withdrawal.

\textbf{\infra \proto Deposit phase steps:}
\begin{enumerate}
  \item The user submits its commitment and the User Obfuscated Data into the dApp SC on the source chain. The dApp SC applies its own business logic to validate a transaction and it sends the user data to the Router SC. The Router SC checks the address of the dApp SC and it emits an event. This event is composed of: commitment, Trustless Public commitment, and the Source Chain ID.
  \item The off-chain entities Oracles capture the event.
  \item The Oracles send the event into the Mixer SC which is located on the Multiplexer Blockchain. The Mixer SC checks that the commitment hasn't been already added. Afterwards, it creates the Merkle Tree leaf, which is the addition of the event parameters. Afterwards, it passes the Merkle Tree Leaf to the Merkle Tree SC. In parallel, the Oracles periodically update the Merkle Tree Root into all the Router SCs.
  \item The dApp Owner nodes checks which Merkle Tree Leaves are from its own dApp, it checks that it has correctly originated on the Source Chain. And through an off-chain process it signs its own Merkle Tree Leaf.
  \item The dApp owner nodes submit the signature to the Mixer SC. This can be upgraded through an off chain process where the user gets the signature while preserving the privacy and anonymity. 
\end{enumerate}

\textbf{\infra \proto Withdrawal phase steps:}
\begin{enumerate}
  \setcounter{enumi}{5}
  \item The user obtains the Merkle Tree Leaf signature from the Mixer SC.
  \item The user creates the Settlement-ZKP.
  \item The user sends the proof plus the agnostic payload data to the Router SC on the destination chain. Optionally, the user can route its message through the Oracles or Relayers.
  \item The Router SC checks the assertions of the ZKP public signals.
  \item The Router SC checks the validity of the Settlement-ZKP via an interface that calls the Verifier-Settlement SC.
  \item The Router SC calls the destination dApp SC and settles the transaction. 
\end{enumerate}

\subsection{Revert Transaction Flow}
Transactions on the destination chain can fail. For example, a user deposits 100 ETH on the source chain but the withdrawal transaction failed on the destination chain. Therefore, the \infra system through the \proto protocol permits the users to have custody of their funds. In other words, the users can revert the transaction and recover the funds, independent from the liveliness of the Oracles and the dApp.

\begin{figure}[H]
\centering
  \includegraphics[width=0.999\textwidth]{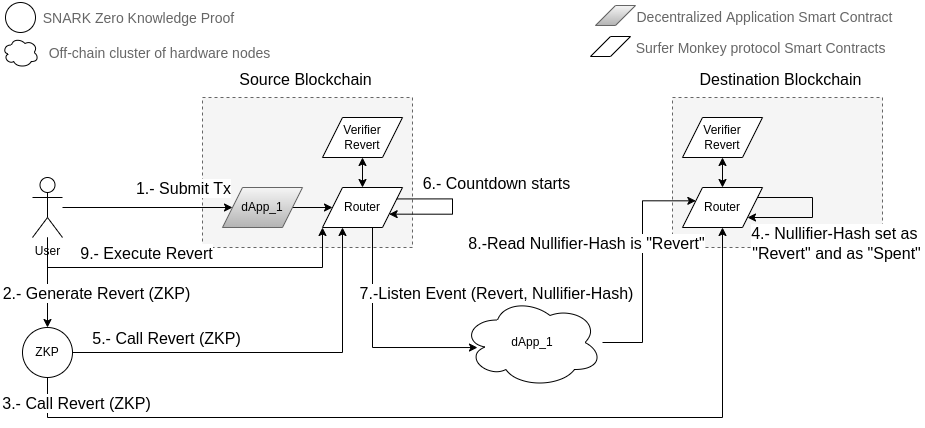}
  \caption{\infra: Revert transaction}
  \label{fig:cancelFlow}
\end{figure}

The revert function can only be called on the correct source chain ID, and it implements a time window. The correct chain ID is verified via the ZKP. And during the time window, the dApp can halt the Revert transaction, if the dApp does not halt the Revert transaction, then the Revert transaction gets executed. The proposed Revert function comes at the cost of breaking the user privacy. The flow from a cancel transaction is depicted in \cref{fig:cancelFlow}.

\textbf{\infra \proto Revert steps:}
\begin{enumerate}
  \item The user submits a transaction to a dApp, for example to deposit 100 ETH.
  \item The user creates a Revert-ZKP.
  \item The user calls the destination Router SC with the Revert-ZKP as a parameter.
  \item The destination Router SC performs the ZKP validity assertion, Merkle Tree checks, destination chain ID checks, and that the transaction has not been spent. Then, the Nullifier-Hash is placed as Spent and as Reverted.
  \item The user calls the source Router SC with the Revert-ZKP as a parameter. The source Router SC checks ZKP validity, the Merkle Tree checks, the source chain ID checks, and that the commitment is in the mapping. 
  \item If the source Router SC checks are OK, the time window countdown starts and a Revert Event is broadcasted.
  \item The Revert Event is captured by the dApp. 
  \item The dApp reads on the destination chain that the Nullifier-Hash has been set as Reverted and as Spent. If true, the dApp does nothing. If false, the dApp halts the Revert transaction on the source Router SC.
  \item Once the countdown is expired, the user can call Execute Revert. The router SC will invoke the Revert function on the dApp, then the dApp SC transfers the funds (100 ETH) back to the address associated to it.
\end{enumerate}

There is no logic into implementing a ZKP that keeps the user Merkle Tree Leaf, commitment, and source chain ID private. As even if we can keep those signals private, and somehow verify them against public data on the source Router SC, the off-chain graph analysis will link both transfers which will relate the source chain with the destination chain. Therefore, there is no reason to implement a more private circuit since the off-chain analysis will break the user privacy even with.

The Revert function can be called by any entity, but they have to submit the Revert-ZKP which is different than the Settlement-ZKP. With this, we achieve three security principles:

\begin{enumerate}
  \item The Oracles cannot arbitrary revert commitments on the source chain, as they cannot reconstruct the Merkle Tree Leaf, and thus the ZKP.
  \item The Oracles cannot take a \proto Settlement-ZKP and perform a revert on the source chain.
  \item Users and Oracles cannot perform a double spent attack as the Revert function has a time window. During the time window, the dApp ensures that the user has indeed flagged the transaction as reverted on the destination chain. The omission of the Revert flag, or the transaction flagged as Spent, will motivate the dApp to halt the transaction.
\end{enumerate}

\subsection{Actors Authentication}
We implement different authentication mechanisms depending on the Actor (\cref{tab:actorsAuth}). We have Private and Public Key pair schemes, Distributed Signature Schemes, and the Proof of Membership. The latter authentication system is verified via the valid \proto ZKP and its public signals.

\begin{table}[tb]
	\centering
    \caption{\infra actors authentication systems.}
    \label{tab:actorsAuth}
    %\vspace{-6pt}
    \begin{tabularx}{.9\textwidth}{cccc}
        \toprule
        \bf Users & \bf Oracles & \bf dApp single owner & \bf dApp distributed ownership\\
        \midrule
        \multirow{1}{*}Proof of & Distributed & 1.- EVM key pair & 1.- EVM key pair shares \\
        membership & signature scheme & 2.- ECDSA key pair & 2.- ECDSA key pair shares\\
        \bottomrule
    \end{tabularx}
\end{table}

\subsubsection{Users}
They are authenticated via the Proof of Membership, thus they are able to construct the correct ZKP with the right public signals. Here, the users prove that they can reconstruct the Merkle Tree Root from the Merkle Tree (Membership) without revealing their Merkle Tree Leaf (Member).

The end users are the actors that hold the private structures of Secret, Nullifier, and the Salt. Therefore, they are the only ones that can recreate the \proto Commitment and the \proto User Obfuscated data which prevents everyone else to reconstruct a Merkle Tree Leaf without their private data structures. Therefore, no else can hijack the Proof of Membership from the Merkle Tee.

\subsubsection{Oracles}
The Oracles are authenticated via a Distributed Signature Scheme for decentralization. Therefore, we implement a distributed Oracle Network from a third party supplier at the cost making the integration to their end point.

Implementing a third party distributed Oracle Network does not implicitly create risks for our system for two reasons. First, the Oracles can inject forged Merkle Tree Roots but they cannot settle forged messages as they cannot reconstruct the Proof of Membership via the ZKP without the dApp signature. Second, the user has custody of their transaction so the user can still make the Revert-ZKP and recover its funds without the need of the Oracle network.

\subsubsection{dApp with a single owner}
In the case of a single entity being the owner of the dApp, the authentication on the SC Mixer for the dApp SC Registry is given by the address message sender that initializes the SC Registry.

As for the dApp Merkle Tree Signature, we have to implement signatures that are supported by the IDEN3 Circom circuit, which are the ECDSA key pairs. If we have a single owner of the dApp, the private Key is the custody from the dApp. 

\subsubsection{dApp with a distributed ownership}
A dApp can be more resilient by implementing a Distributed Signature Scheme. For the case of the dApp Registry, the dApp can distribute the Private Key into multiple shares among the Nodes and the reconstruction of the threshold shares will output the Private Key from the address that calls the Registry SC function.

In the same way, for the use case of the Merkle Tree Leaf signature, the dApp can distribute its ECDSA Private Keys into multiple shares which can be reconstructed using a threshold of shares.

\section{System Resilience and Analysis}
We depict several attacks that can be done against the \infra system and describe how \infra resists against them. We considered various edge cases during the design and implementation phases.

\subsection{Off-chain Attacks and Denial of Service}
Off-chain attacks are defined as attacks performed outside the Blockchain environment, directly targeting to the Blockchain nodes such as Oracle nodes and dApp nodes. As our system implements a decentralized proactive security scheme, the attacker must get control of the Oracle threshold nodes and control of each dApp node cluster where the attackers want to strike.

For example, if the attackers want to perform forged settlements on two dApps, the attackers must take over the Oracle Network threshold nodes, the dApp\_1 threshold nodes, and also the dApp\_2 threshold nodes. This increases the security bar in \infra against cyberattacks. However, \infra is not immune and each Node owner must perform their due diligence to keep the hardware node and the private keys data secure.

Furthermore, even in the case of a Denial of Service on the Oracles and on the dApps, the users can still revert their transactions and obtain their funds back.

\subsection{Oracle Attacks}
The attacks that the Oracles can do to the network fall into two main categories: Network abuse and Censorship.

Th Oracles must carry data from one Blockchain and settle it in another Blockchain. They perform this task through a consensus mechanism. The Oracles Consensus mechanism falls out of scope of the \infra protocol thanks to the implementation of the \infra decentralized proactive security.

\subsubsection{Network Abuse}
We define a Network Abuse as a Game Theory move to obtain a higher reward by the Rational Players (Oracles). Per se, this is not an attack as the Oracles have the keys to settle transactions.

This move is initiated by the Oracles that inject forged Merkle Tree Roots in the supported Blockchains. Second, the Oracles create a valid ZKP whose public signals match the forged Merkle Tree Root. With this, the Oracles are able to settle forged transactions and steal funds from the dApps.

Our solution falls in the scope of proactive cybersecurity. We decentralize our solution by giving each dApp the choice to sign or not a specific Merkle Tree Leaf. The creation of the ZKP only works if the dApps have signed their own Merkle Tree Leaves. This signature is required by the user to reconstruct the Membership Proof through the \acp{zksnark}, but without revealing its Member (Merkle Tree Leaf). All in all, this signature avoids the Oracles from been able to reconstruct forged settlements, as they do not have the dApp signature.

In order for the dApp users to have certainty that the dApp won't align with the evil Oracles, the dApp can implement a Distributed Signature Scheme where none of the signers has a rational sentiment to align with the Oracles. Going further, each Node signer has a rational motivation for the dApp success.

\subsubsection{Censorship}
The Oracles can decide to censor a User. This can be because of the source user address, the destination user address, or the payload that is in transit does not align with the Oracles motivations. As An example, the Oracles can censor a user vote for a Decentralized Autonomous Organization (DAO), or a specific DeFi swap trade which perpetually locks the user funds on the source chain.

First, we leverage the ZKP privacy principles. Through the \proto messages we provide user privacy, destination blockchain anonymity, and payload data obfuscation. Furthermore, when the user submits the ZKP into the destination Blockchain, the Oracles cannot link this transaction to the source transaction.

Second, the \proto protocol allows the user to revert the transaction before it gets settled on the destination Blockchain. This process is independent from the Oracles or the dApps participation. Therefore, the users can still revert the transaction on the source chain and obtain their funds back even if the Oracles censor arbitrary transactions.

\subsection{dApp Attacks}
The dApps have three main functions. First, submit the dApp Global Hash. Second, sign user Merkle Tree Leaves. Third, flag forged revert transactions.

\textbf{dApp attack vectors:} 
\begin{enumerate}
  \item dApp Global Hash: A dApp can try to steal another dApp Global Hash before the legit dApp owner registers the address. Furthermore, if this succeeds, the malicious actor could ask for a ransom to unregister the dApp Global Hash. This attack can be performed via two ways: either through an MEV move, or by wrongfully registering the dApp Global Hash before the legitimate dApp owner does it. 
  \item Sign User Merkle Tree Leaves: A dApp can sign malicious user Merkle Tree Leaves, or they can choose to not sign the correct Merkle Tree Leaves.
  \item Flag forged revert transactions: The attack here comes in two sub-vectors: the dApp can choose to flag correct revert transactions as forged, or choose not to flag forged transactions and thus the forged revert calls will be successfully executed.
\end{enumerate}

\textbf{dApp attack vectors solutions:} 
\begin{enumerate}
  \item dApp Global Hash: The dApp SC has a special interface that is only able to be called by the dApp owner. Through this interface, the dApp SC calls the Router SC registration function and the dApp owner passes the values of all the supported SC addresses, except the address of the SC where the Router is been called from. Then, the Router SC constructs the dApp Hash registry by making the Keccak256-Hash of the supported SC address that the dApp owner sent and the address of the SC where it has been called from. This solution counters MEV and is also a long term resilience for hijacking.
  \item Signs Users Merkle Tree Leaves: In order to have an optimal performance from a Rational Players Game Theory point of view and from cybersecurity standards, the dApp must implement a Distributed Signature Scheme where each of the signing nodes has no rational motivations to align with malicious actors.
  \item Flag forged revert transaction: Same as the previous item.
\end{enumerate}

\subsection{User Attacks}
The users will try to game the system which they can do through the transaction data (ZKP), and within the logic inside the deposit, withdraw and revert function.

The user moves and attacks consists of:
\begin{enumerate}
  \item Double spending.
  \item Double spending through the Settlement-Revert logic.
  \item Withdrawal in the wrong chain ID.
  \item Set a different payload data in the withdraw transaction phase.
  \item Call a wrong dApp.
  \item Change sensitive data.
\end{enumerate}

\subsubsection{Invalid ZKP}
Once the user has created a valid ZKP, it cannot change any of the proof nor public signals parameters. As such, changes will lead to the verification during the Withdrawal function to output false and to a failed withdrawal.

\subsubsection{Evil ZKP public signals}
The user can create a valid ZKP with forged public signals. Therefore the \proto protocol has several constraints:

\begin{enumerate}
  \item Double spending: Avoided through the Nullifier-Hash.
  \item Double spending through a Settlement-Revert logic: Avoided through chained actions. First the user must set the transaction as Reverted on the destination chain; it is only then that it can call the Withdraw function. If the user tries to cheat, the rational dApp nodes from the Distributed Signature Scheme will abort it. Furthermore, we have two different ZKPs for the Proof of Membership in order to avoid data replication.
  \item Withdrawal with the wrong chain ID: A parameter of the Merkle Tree Leaf consists of the destination chain ID, therefore it cannot be changed.
  \item Revert with the wrong chain ID: A parameter of the Merkle Tree Leaf consists of the source chain ID, therefore it cannot be changed. Furthermore, this parameter is set by the Router SC.
  \item Set a different payload data in the withdraw transaction phase: A parameter of the Merkle Tree Leaf consists of the payload, therefore it cannot be changed.
  \item Call a wrong dApp: A parameter of the Merkle Tree Leaf consists of the dApp Global Hash, therefore it cannot be changed. Furthermore, the Router SC is the one that places the dApp Global Hash.
  \item Change the sensitive data: A parameter of the Merkle Tree Leaf consists of the sensitive data, therefore it cannot be changed. Furthermore, this data is input by the Router SC.
\end{enumerate}

In summary, if the user changes any of the input signals while creating the ZKP, then the user won't be able to reconstruct the correct Merkle Tree Root. And therefore, it won't be able to deliver a valid Proof of Membership.

\subsection{MEV Background}
As of July 2022, the aggregated extracted Maximal Exctractable Value (MEV) from the users is \$663,716,529; from that value the cumulative MEV miner payment is of \$237.2 millions~\cite{flashbots}.

The MEV is a new concept that first came out in 2019 as ``Miner Extractable Value''~\cite{flashBoys}. MEV leverages the Blockchain transparency principle to search for transactions in the mempool, where the user transaction negotiation parameters are transparent, in order to extract as much possible value from the user. The MEV extraction is performed independently of the intent of the user.

MEV moves can be performed on any system where users transactions can be reordered and where the MEV Players can see the transparent payload data from the transaction. We have already covered the Oracles power and Censorship rights that they have within the Blockchain intercommunication systems. We have also set system boundaries so that the Oracles cannot perform MEV moves or Censorship on the \infra system. Therefore, let's focus on the MEV from the Rational Players defined as standards Users (MEV Searchers) and Blockchain Miners.

A Blockchain Miner MEV move relies on the transaction ordering to achieve their MEV target. The Blockchain Miners can do a re-ordering of the Blocks, transaction insertion, or censor a user transaction. On the other hand, the users can Front-Run, Back-run or Sandwich the Rival transaction.

\subsubsection{Front-run, back-run and sandwich}
Here we give an overall view on the tree terms of Front-Run, Back-Run and Sandwich game plays.

The Front-Run move is performed via a Bot that is monitoring the Blockchain mempool. Once a profitable transaction is spotted, the MEV Player will clone the Rival transaction and change the Rival address with the Player address. After running locally the new transaction to check that the move is indeed profitable, the Player submits its transaction to the Blockchain mempool and pay a higher Gas Price to have its transaction executed before the Rival transaction~\cite{ethMev}.

On the other hand, the Back-Run move consist on taking advantage from the surplus transaction created by another user, and thus the MEV Player places the transaction just behind the user.

Finally, the Sandwich move is created with big user trading transactions as large transaction will impact the tokens pair price. Basically, the Sandwich move consists on placing two transactions, one just before the Rival transaction, and one just after it. The Front-Running transaction buys the same token at a cheaper price just before the Rival transaction gets executed. After the Rival transaction gets executed, the token pair price goes up. Here, the MEV Player sells the tokens at a higher price through the Back-Run transaction~\cite{sandwich}.

\subsection{\infra MEV Resilience}
The users data is obfuscated within the \proto protocol. Therefore, no Blockchain Miner, Oracle, or User can tell what the payload data and the destination Blockchain are. This shields the users against MEV moves for cross-chain communication: on the source Blockchain, during transit, and optionally on the destination chain.

The reason why the MEV opportunity gap in the destination chain still persists is because the user reveals the public signals of the ZKP in the destination chain. Therefore, we require to integrate \infra with another MEV single-chain resilience protocol on the destination chain, such as a commit-reveal mechanism with a time frame gap in between, or a Flashbots private mempool~\cite{ethMev}.

\subsection{Privacy}
The whole point of the \proto protocol implemented in \infra is to deliver privacy and anonymity to its users. In order for the users to increase their security bar, they must wait a certain time before they reveal their ZKP public signals on the destination chain.

Furthermore, they must wait until there is a sufficient amount of Deposit transactions into the system so their Withdraw transaction linkage gets obfuscated. We have to highlight that the graph analysis is reduced from the whole system to a dApp graph level because at this moment we are only allowing same dApp to dApp communication. However, this can be expanded through a dApp that advances transactions to any any other dApp and that has its own Liquidity Pools.

\subsubsection{Compromised transactions due to heuristics}
As shown in the Tutela open source project, there are transactions which privacy has been compromised through careless user behaviour, even though the users used a ZKP transaction Mixer such as \tornado~\cite{tutela}.

The Tutela open source project focuses on privacy compromised on a single chain environment. We are working on a multi-chain system, therefore we have the same heuristics from single-chain environment plus heuristics from multi-chain environments yet to be researched.

The Tutela project heuristics found in single chain environments are:
\begin{enumerate}
  \item Address Match: Same address in the Deposit and in the Withdrawal.
  \item Unique Gas Price: A specific user tends to choose the same gas price.
  \item Linked Blockchain addresses: The deposit and withdrawal addresses are linked through the Blockchain graph because both addresses have interacted before or after utilizing the Blockchain mixer.
  \item Multiple Denomination: Addresses whose portfolio of deposit transactions are identical to the withdraw addresses portfolio.
\end{enumerate}

\subsection{System Constraints}
The system is constraint to certain user actions, dApp and Oracles requirements. Here we highlight them:
\begin{enumerate}
  \item The dApp must sign each leaf.
  \item In case of a Revert call, the dApp must analyze the Nullifier-Hash in the destination chain.
  \item The users can only perform one action: Settle or Revert. They cannot change decisions.
  \item The dApp Distributed Nodes have no rational intentions to align with evil Oracles or other evil entities.
  \item The users have custody of their Secret, Nullifier and Salt, and they must not loose these numbers. If they loose them, they will loose access to their funds or actions.
  \item The dApp must update its dApp Global Hash in all the Router SCs.
  \item The dApp cannot speak with another dApp. That can be done as an application level, not at a \proto protocol level.
\end{enumerate}

\section{Proof of Concept Implementation}
The implementation process has been done in three main steps. This has allowed us to tackle the problem from the most fundamental part up to the application level.

Implementation process:
\begin{enumerate}
  \item Understand \acp{zksnark} from an application point of view. Here we leveraged the Circom language by IDEN3.
  \item Reverse engineering \tornado. Code repository: \textit{tornado-core}~\cite{tornadocashcore}.
  \item We apply the code, system, knowledge, principles and constrains from \tornado into \infra.
\end{enumerate}

We have done the \acf{poc}\glsunset{poc} on a sandbox environment of a local test net. The \acf{ide}\glsunset{ide} that we used was Truffle because we already had experience with this IDE.

\paragraph{Blockchains.}
To minimize efforts, the PoC was done on a single Ethereum Virtual Machine (EVM) Blockchain since if the system works in one EVM Blockchain, it will work in the rest of EVM Blockchains. The constraint to make the PoC with an EVM Blockchain comes from the Circom language. As the export function from Circom delivers an SC written in Solidity which is compatible with the EVM Blockchains. 

We require to deliver a \ac{zksnark} verification inside the block computational steps limits. Therefore, for a broader implementation with no EVM compatible chains, we would have to implement the verification of Elliptic Curves operations within the pre-compiled SC for BN128 or BLS12-381 curves. Ethereum already supports those operations on the EIP-196~\cite{eip196} and EIP-197~\cite{eip197}. 

\paragraph{Authentication}
For the PoC we have implemented a single entity for both owners: dApp owner and Oracle network. Therefore, the dApp and Oracle authentication checks are performed by a Modifier function on the SC that checks the address from the message sender. 

\paragraph{Merkle tree SC}
The Merkle Tree has a range of depth levels from 1 to 32. With 32 levels, we will get $ 2^{32} = 4,294,967,296 $ Leaves, but it will be more gas costly to input a Leaf into the Merkle Tree. Therefore, we have to find the right balance on the Merkle Tree Depth. Our system has two main steps that are affected by the Merkle Tree Depth: those are to inject the Merkle Tree Leaf, and check the validity of the ZKP in the Verifier SC.

\paragraph{dApp SC}
We have implemented a dApp, that is a simple Hello World style program, where the user sends a string on the source dApp and this string is transmitted to the destination chain.

\subsection{Reverse engineering \tornado}
In order to reverse engineer \tornado, we had to breakdown \tornado into multiple steps.
\tornado reverse engineer phases:
\begin{enumerate}
  \item Understand \tornado ZKP circuit: \textit{Withdraw}~\cite{tornadocashcore}.
  \item Test the circuit through command line. For this, we have to obtain a valid input JSON file to test the circuit.
  \item Create a dummy \infra ZKP circuit and test it.
  \item Understand \tornado JS minimal demo: \textit{minimal-demo.js}~\cite{tornadocashcore}.
  \item Create a local JS to test the creation of the \infra ZKP via JS, and test the validity of the proof through the JS.
  \item Understand the \tornado SCs: \textit{Tornado.sol, ETHTornado.sol, MerkleTreeWithHistory.sol}~\cite{tornadocashcore}.
  \item Create a dummy \infra SC.
  \item Understand \tornado interaction of the JS IDEN3 libraries with the SCs: \textit{ETHTornado.test.js}~\cite{tornadocashcore}.
  \item Implement a dummy interaction of the \infra SCs with the JS client so the user can create a \ac{zksnark} and send it to the Blockchain SC.
  \item Improve the \infra design, SCs and scripts until our system requirements are met. 
\end{enumerate}

\subsection{Complications and Learnings}
The main complication in the development of the project has been on the integration of 3 compilers: Circom, JavaScript (JS), and Solidity. This means that each compiler has a deterministic behaviour with specific API libraries. For the JavaScript environment we have implemented the EthersJS API. All in all, we require assurance that the three compilers can see the same truth, if not, the user will not be able to reconstruct the Proof of Membership via the \acp{zksnark}, and the user funds will be perpetually locked.

Other minor complications were on the construction of a valid input JSON file. As well as understanding the array system within the Circom language.

\begin{figure}[H]
\centering
  \includegraphics[width=0.7\textwidth]{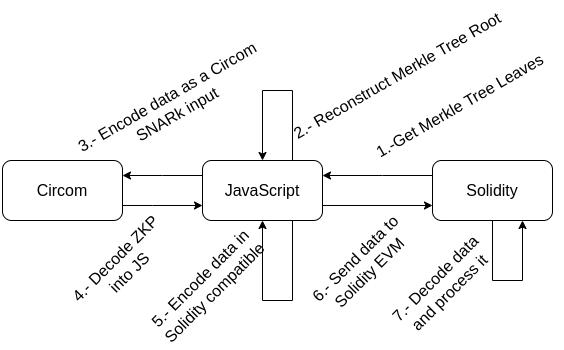}
  \caption{\infra: Integration of languages and steps to reconstruct the Proof of Membership}
  \label{fig:integration}
\end{figure}

In \cref{fig:integration}, we show the compiler integration path to reconstruct the Proof of Membership. Steps and challenges:
\begin{enumerate}
  \item Obtain all the Merkle Tree leaves from the Merkle Tree SC. 
  \item We have to clean the data. Afterwards, we shall reconstruct the Merkle Tree and the Merkle Tree Path to the user Merkle Tree Leaf.
  \item Now, we have to encode all the input signals of the ZKP circuit from JS into Circom compatible inputs. Furthermore, we require that the Pedersen-Hash that the IDEN3 JS-API produces is equal to the IDEN3 Circom compiler.
  \item Once we obtain the ZKP, we have to decode the public signals so we can clean them and perform data transformation on them.
  \item After the ZKP public signals have been cleaned, we can encode the public signals and the proof into an EVM transaction that matches the expected data types from the Verifier SC. Highlight: The Verifier SC is an export from the Circuit after the initialization ceremony. This export is done via Circom.
  \item We send this data to the EVM mempool.
  \item The transaction gets executed on the SC and the values are decoded from the encapsulate transaction data into individual data that is sent to the Verifier SC.
\end{enumerate}

Even though this process has been mostly cleared by \tornado and IDEN3, it is of vital importance that we (engineers) understand it. The reason is that, in \infra, we had to perform several updates from \tornado which was developed with Circom 1.0.0 to \infra (Circom 2.0.0). This lead us to complications in the encoding of the ZKP to make it compatible with Solidity as well as with older versions of the JavaScript API libraries with the Circom 2.0.0 version and the Solidity 0.7.0 version.

\section{Evaluation}
The experiments environment has been conducted in a preliminary sandbox local EVM Blockchain network via Truffle. We have performed unit testing from each function to address the correct behaviour at a function level. Afterwards, we tested an End to End implementation where we assessed the overall behaviour of \infra. In order to meet the project deadline, we have done the experiments from the PoC with a simpler ZKP circuit that does not integrate the signature verification circuit.

Apart from testing the successfully execution of the functions, we also tested the failing scenarios. All in all, we obtained a preliminary result of how the system behaves on favourable and adverse scenarios.

\subsection{Unit Tests}
We have executed unit testing at the levels of the Circuit, JS, and the Solidity SCs. We have over 47 unit tests. They range from being independent of the current EVM state to unit testing that is EVM state dependant. The latter tests require to have an initial setup, we performed the setup as a pyramidal development of unit testing where the function is tested each time it is called.

\subsection{System Tests}
We have coded a test scenario for an end to end flow in order to assess the perform from the overall system. This has been done to test the integration of all the unit tests as well as to track the changes on the EVM state machine.

\subsection{Gas Execution Cost}
The motivation to find the gas costs from our system relies on optimizing its usability. As the Mixer SC can support as many transactions as the Merkle Tree SC can hash. Therefore, we must research the optimal Merkle Tree Depth point.

From a gas consumption point of view, the system has three main steps: the \proto message submit, the injection of the Merkle Tree Leaf into the Mixer SC, and the \proto message settlement on the destination chain. From these steps, only the injection of the Merkle Tree Leaf and the \proto message settlement are affected by the Merkle Tree Depth. Therefore, we have concentrated our experiments on gas execution cost to asses the Merkle Tree Depth. In the next figures, we have the evolution of the Gas execution cost against the Merkle Tree Depth for the Merkle Tree leaf injection and the \proto Withdraw phase.

\begin{figure}[tb]
\centering
  \includegraphics[width=0.8\textwidth]{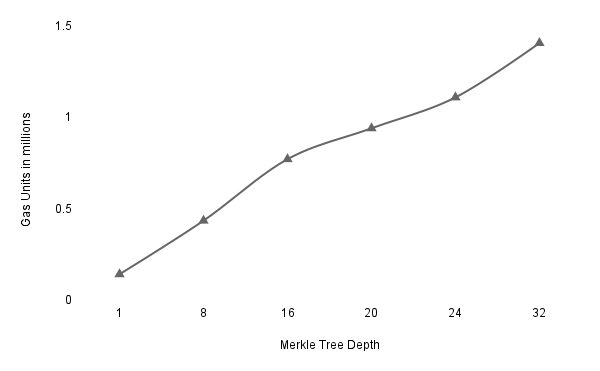}
  \caption{Leaf injection into the Merkle Tree: Gas evolution against the Merkle Tree Depth}
  \label{fig:gasDactLeafInject}
\end{figure}

In \cref{fig:gasDactLeafInject}, the graph shows the evolution of the curve of the gas cost for the injection of the Merkle Tree Leaf into the Merkle Tree. Here is where we have the greatest impact in terms of gas expenditure due to the Merkle Tree SC having a ``for loop'' in the ``insert'' function. This function loops from the depth 0 till the maximum depth, in order to obtain the hash from the current Merkle Tree Level.

\begin{figure}[tb]
\centering
  \includegraphics[width=0.8\textwidth]{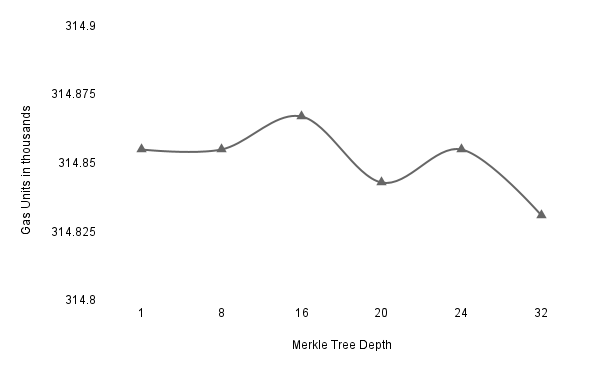}
  \caption{\proto Withdraw phase: Gas evolution against the Merkle Tree Depth}
  \label{fig:gasDactWithdraw}
\end{figure}

In \cref{fig:gasDactWithdraw}, the curve is on the \proto Withdraw phase that includes SNARK verification. There isn't any significant change on the curve as more Merkle Tree Depths are added. We do note fluctuations with a lower tendency but they are too small to be considered meaningful. This is because the SNARKs have the principles of succinct and sublinear so the proofs are verified using similar amounts of time and that more complex \ac{zksnark} circuits will be roughly a little bit more costly to be verified. 

\begin{figure}[tb]
\centering
  \includegraphics[width=0.8\textwidth]{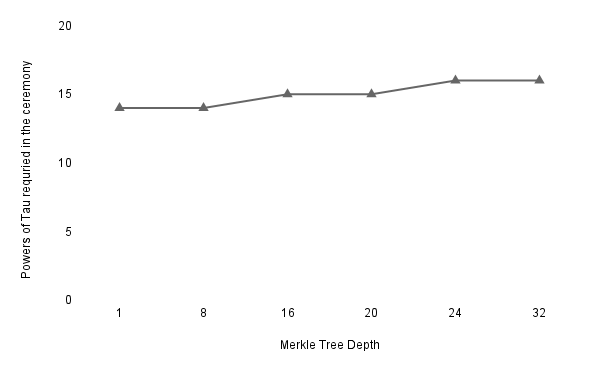}
  \caption{Powers of Tau evolution against circuit complexity}
  \label{fig:powerTau}
\end{figure}

In \cref{fig:powerTau} we have the evolution of the Powers of Tau against the complexity of the circuit. The Powers of Tau ceremony increases as the number of constraints and computational requirements from the circuit increases. Therefore, as we have more Merkle Tree levels, we also need to have a longer input signal array.

We have made practical experiments with the SNARK principles of succinct and sublinear. Even though the circuit becomes more complex as the Merkle Tree Depth increases, the computational expense to verify the ZKP remains fairly constant. 
Finally here is the summary of the experiments: 
\begin{enumerate}
  \item The Leaf injection into the Merkle Tree consumes a big portion of gas, furthermore, it consumes more gas as more Merkle Tree Depth we have.
  \item We have some fluctuations on the gas expenditure for the \proto Withdraw phase that we can use to leverage the system. However, they do not deliver substantial changes thanks to the stability of the verification time of SNARKs.
  \item Even though the Power of Tau increased following the Merkle Tree Depth, this did not had a big impact for the proof verification in terms of gas cost thanks to the SNARKs properties of being succinct and sublinear.
\end{enumerate}

In these experiments we have verified the SNARK properties in terms of being succinct and sublinear. On the other hand, a Merkle Tree Depth candidate can be the level 20 as it is on a good trade-off among the gas consumption and Merkle Tree usability. A stronger solution will be to deploy the Merkle Tree SC within the Mixer SC, at a maximum depth of 32, on an EVM Layer-2 solution.

\section{Design Decisions}
We mention the must prominent design decisions of \proto and \infra.

\begin{enumerate}
  \item \textbf{Multiplexer blockchain:} A Multiplexer Blockchain allows us to concentrate all transactions into a single Mixer SC. This SC creates a transaction Mixer by storing transactions from many Blockchains in a single Merkle Tree. When it changes, the Merkle Tree Root is updated in all the supported Blockchains. With this, we can have a single source of truth which is replicated on the other Blockchains. Furthermore, by having the users making the \proto Withdraw phase directly on the destination chain and not on the Multiplexer Blockchain, we minimize the MEV opportunity points.
  
  \item \textbf{dApp signature:} To avoid the Oracles from abusing the network by injecting forged Merkle Tree Roots into the Router SCs. There is no motivation for the Oracle to misbehave, if later on they cannot perform a settlement because they do not have the dApp signature for their evil transaction.
  
  \item \textbf{Merkle tree leaf data structures:} In the Merkle Tree Leaf, there are two main types of data structures: data that the user requires to transact, and data that the system requires to be secure. The data that the system requires to be secured is input by the \proto SCs.
  
  \item \textbf{Data structures integer tiers:} In the \proto Structures, we have different tier ranges for the numbers of the Chain ID and the version. This is to avoid some malevolence action through data collision. 
  
  \item \textbf{Global hash dApp:} It has two objectives: Security and Privacy.
  
  From a Security point of view, with our current design the dApp SCs require to give permission to be called by the Router SC. And the Router SC will call the dApp to settle a transaction if both the ZKP public signals are correct and the ZKP is valid. This means that an evil user could create a dApp\_1, and settle a transaction on dApp\_2 without paying dApp\_2 on the source chain.
  
  From a Privacy point of view, the source Blockchain SC cannot use the destination SC address as this information will directly link the transaction on the source chain to the destination chain.
  
  Therefore, the dApp Global Hash works as a unique global alias that also shields the dApp from calls by unauthorized dApps. We can have a dApp calling another dApp, but this must be managed at an application level, not at a \proto protocol level.
  
  \item \textbf{Obfuscate user data with a Salt number:} To maintain user privacy on their payload within a Revert call. If we were using the Nullifier-Hash to obfuscate the user intentions, then the user intentions will also be revealed in case of a Revert call.
  
  \item \textbf{Revert functionality reactive system:} The \proto protocol Withdraw phase follows a strict proactive security system. However, we introduce a reactive system for the \proto Revert function to remove any active requirement of liveliness from the dApp nodes and the Oracles nodes. The users can always Revert their transaction and thus keep custody of their funds. Furthermore, it also motivates the dApp to be vigilant and do not become lazy.
  
  \item \textbf{dApp resilience rules on the Revert function:} This is a safeguard mechanism for the dApps. If for any reason, the dApp is not able to flag the revert forged transactions, then at least it can win some time to perform emergency operations. For example, these salvage operations can give time to the Liquidity Providers to withdraw their liquidity during a total annihilation of the dApp Nodes.
\end{enumerate}

\section{Future Research and Development}
Through the \proto protocol, we have set the foundations for an anonymous Blockchain intercommunication system. The PoC has given initial positive results on how the system behaves in favourable circumstances and in adverse scenarios. However, in order to improve the current \proto protocol version towards a full fledged vision (\proto protocol version 2) further research is required.

\subsection{Accumulators}
In the current implementation of the \proto messages, in order to create the Proof of Membership we use Merkle Trees which Root must be recomputed in the \ac{zksnark} circuit. The current implementation of the Merkle Tree has a constraint of a maximum depth level, which impacts the number of transaction that the system can support. And it brings complexity into the ZKP and into the SC.

Therefore, we could leverage efficient cryptographic Accumulators as a one way membership hash function~\cite{batching-rsa-accumulators}. With this, we can give a more efficient implementation to the candidates to prove their member status. 

\subsection{Distributed Key Generation}
To achieve a higher cybersecurity resilience and a stronger Rational Players honest intentions, we motivate dApps to implement Distributed Signature Schemes.

However, one of the drawbacks of Distributed Signature Schemes is that the public key shall change every time there is an update on the network configuration. This means, that each time a new node enters of leaves the dApp network, the public key shall be updated with a new key pair. This brings security and managements constrains.

Therefore, the research on Distributed Key Generation with a constant Public Key is a factor that makes the managing of the dApp network and security simpler~\cite{dkg-internet}. This can be done through the distribution of shares from the Private Key, and in parallel apply a membership mechanism to allow new entities to place their shares and ban expelled entities from entering the system boundary. 

\subsection{Blockchain Reorganisations}
Blockchain blocks reorganisation (Reorgs), is a security constraint that exists in every Blockchain with a probabilistic finality. This means, that users shall wait a certain amount of blocks for their transaction to be consider final. The more blocks they wait, the higher the probability that the block where the transaction has been inserted won't be re-organized and, thus loose the finality.

At this stage, the most secure thing we can do to avoid the reorgs problem from probabilistic Blockchains is to wait longer before the users input their transaction on the \proto Withdraw phase.

All in all, we must perform deeper research on this realm, to see if we can overcome the trade off between security and user experience.

\subsection{Compromised Intercommunication Blockchain Transactions due to Heuristics}
The Tutela whitepaper~\cite{tutela} has mentioned some careless user behaviour which led to a compromised anonymity on their transaction through the \tornado single-chain mixer, thus breaking the user privacy.

Apart from applying single-chain privacy mixer rules, we must research new rules that the users shall apply in order to keep their privacy on a multi Blockchain mixer ecosystem. 

\subsection{Minimize Gas Transaction}
In order to reach mass adoption, we have to minimize the gas transaction cost from the \proto messages within the \infra system.

\section{Conclusion}
We provide a solution for the anonymity of Blockchain intercommunication. Our first contribution is the \acf{dact} protocol that leverages Proofs of Membership via ZKPs. Our second contribution is the \infra system which is the decentralized infrastructure that supports the \proto messages.

Through the \infra system, the users can have an end to end anonymity in a multi Blockchain environment as no party can reconstruct the link between the input and output transaction. We achieve this by breaking the on-chain analysis through the ZKPs, and the off-chain analysis through a multiplexer Blockchain where we concentrate all the user transactions into a Mixer SC.

\infra implements a decentralized proactive security system. The dApp owner signs the user leaves, this signature is later on used by the users to reconstruct the ZKP. The ZK circuit is designed to maintain the privacy, and therefore it does not reveal the signature nor the message. Furthermore, the dApp can opt to implement a Distributed Signature Scheme where each node signer has an honest motivation for the success of the dApp. With this, we raise the security bar against cyberattacks and shield the system against Rational Players (selfish entities that always look for their best interests).

\infra supports a Revert function so that the users can have full custody of their funds as reverting does not require an active participation from the Oracles nor the dApp nodes. We implement a reactive optimistic security principle in the Revert function for two reasons. First, we want the users to be able to recover their funds without depending on any entity. Second, to motivate the dApp to remain vigilant.

Our system is secure by default~\cite{secureByDefault}. We have designed and implemented a solution where the users are not able to use the system in an insecured way.
First, we prioritized security, then usability.

For the PoC, we have leveraged the audited codes from \tornado and implemented the ZKP IDEN3 libraries. We have done this process by reverse engineering \tornado. The PoC is done using 3 programming languages: Circom, JS and Solidity. A challenge of the PoC was that all the programming languages from the system needed to reconstruct the same truth (Merkle Tree). If not, the users funds can be perpetually locked.

We have obtained preliminary positive results from the experiments. We have observed the behaviour of the system for both the successful and the adverse scenarios. The gas cost experiments have verified the SNARKs principles of being succinct and sublinear. The SNARK proof verification gas cost remains quite constant even with a more complex circuit. On the other hand, the depth of the Merkle Tree has a big impact on the gas costs to submit a Leaf.

The \proto protocol does not allow for interaction between different dApps. However, through the \infra system we can have communication among different dApps. This can be done as a dApp with Liquidity Pools that covers the expenses of the calls and advances the transaction to the other dApp.

Thanks to the privacy and anonymity of \infra, there is no cross-chain MEV opportunity points. The users cannot be censored since the data is obfuscated and it is not possible to know the user intentions nor the destination. All these factors open a new gate for us to innovate, as we can have private votes withing a DAO, private games, and a new phase for the Anonymous Decentralized Finance (aDeFi).

\bibliographystyle{acm}
\bibliography{bibliography.bib}

\end{document}